\documentclass[twocolumn,pra,superscriptaddress,showpacs,aps]{revtex4-1}
\usepackage{amsmath}
\usepackage{amssymb}
\usepackage{graphicx}
\usepackage{txfonts}
\usepackage{color}

\begin{document}

\title{Rashba-type spin-orbit coupling in bilayer Bose-Einstein condensates}

\author{S.-W. Su}

\affiliation{Department of Physics and Graduate Institute of Photonics, National
Changhua University of Education, Changhua 50058, Taiwan}

\author{S.-C. Gou}

\email{scgou@cc.ncue.edu.tw}

\affiliation{Department of Physics and Graduate Institute of Photonics, National
Changhua University of Education, Changhua 50058, Taiwan}
\affiliation{National Center for Theoretical Sciences, Physics Division, Hsinchu 300, Taiwan}

\author{Q. Sun}

\affiliation{Department of Physics, Capital Normal University, Beijing 100048,
China}

\author{L. Wen}

\affiliation{College of Physics and Electronic Engineering, Chongqing Normal University,
Chongqing, 401331, China}

\author{W.-M. Liu}

\affiliation{Beijing National Laboratory for Condensed Matter Physics, Institute
of Physics, Chinese Academy of Sciences, Beijing 100190, China}

\author{A.-C. Ji}

\email{andrewjee@sina.com}

\affiliation{Department of Physics, Capital Normal University, Beijing 100048,
China}

\author{J. Ruseckas}

\affiliation{Institute of Theoretical Physics and Astronomy, Vilnius University,
Saul\.{e}tekio Ave. 3, Vilnius,10222, Lithuania}

\author{G. Juzeli\={u}nas}

\email{gediminas.juzeliunas@tfai.vu.lt}

\affiliation{Institute of Theoretical Physics and Astronomy, Vilnius University,
Saul\.{e}tekio Ave. 3, Vilnius,10222, Lithuania}

\date{\today{}}

\begin{abstract}
We explore a way of producing the Rashba spin-orbit coupling (SOC)
for ultracold atoms by using a two-component (spinor) atomic
Bose-Einstein condensate (BEC) confined in a bilayer geometry. The
SOC of the Rashba type is created if the atoms pick up a $\pi$ phase
after completing a cyclic transition between four combined spin-layer
states composed of two spin and two layer states. The cyclic
coupling of the spin-layer states is carried out by combining an intralayer
Raman coupling and an interlayer laser assisted tunneling. We theoretically determine the ground-state phases of the spin-orbit-coupled
BEC for various strengths of the atom-atom interaction and the laser-assisted
coupling. It is shown that the bilayer scheme provides a diverse ground-state
phase diagram. In an intermediate range of the atom-light coupling
two interlacing lattices of half-skyrmions and half-antiskyrmions
are spontaneously created. In the strong-coupling regime, where the
SOC of the Rashba-type is formed, the ground state represents plane-wave
or standing-wave phases depending on the interaction between the atoms.
A variational analysis is shown to be in a good agreement with the
numerical results.
\end{abstract}


\maketitle

\section{Introduction}

Following the first realization of artificial (synthetic) magnetic
field for ultracold neutral atoms \cite{Lin2009}, quantum degenerate
gases have provided a highly controllable test bed for studying the
dynamics of quantum systems subjected to gauge potentials \cite{Struck2012,Aidelsburger2013,Miyake2013,Esslinger2014}.
A possible way of creating synthetic gauge potentials for electrically
neutral atoms relies on the adiabatic following of one of the atomic
states ``dressed'' by the atom-light interaction \cite{Dum1996,Visser1998,Juzeliunas2004,Juzeliunas2006,Lin2009,Dalibard2011,Goldman2014}.
Such atoms can experience a light-induced Lorentz-like force, thus
mimicking the dynamics of charged particles in a magnetic field \cite{Juzeliunas2004,Juzeliunas2006,Lin2009,Dalibard2011,Lewenstein2012,Goldman2014}.
Likewise, non-Abelian gauge potentials can be created when a manifold
of degenerate dressed states of atom-light interaction is involved
\cite{Ruseckas2005,Dalibard2011,Lewenstein2012,Goldman2014,Zhai2014-review}.

An important implication of the synthetic non-Abelian gauge potentials
is that they provide a coupling between the center-of-mass motion
and the internal (spin or quasi-spin) degrees of freedom, forming
an effective spin-orbit coupling (SOC). A variety of novel phenomena
has been predicted for such systems, for example, the stripe phase
and vortex structure in the ground states of spin-orbit-coupled Bose-Einstein
condensates (BECs) \cite{Wang2010,Xu2011,Liu2012PRA,Radic2011,Zhou2011,Sinha2011,Ozawa2012a,Ozawa2012b,Liu2013PRA,Chen2014PRA,Han2015,Su2015},
the Rashba pairing bound states (Rashbons) \cite{Jiang2011,Vyasanakere2012}
and topological superfluidity \cite{Zhou2011PRA,Liu2012,Liu2013}
in fermionic gases, as well as the superfluidity and Mott-insulating
phases of spin-orbit-coupled quantum gases in optical lattice \cite{Cai2012,Radic2012,Cole2012,Xu2014,Chen2016}.

The synthetic SOC has been experimentally implemented for boson \cite{Lin2011,Zhang2012}
and fermion \cite{Wang2012,Cheuk2012} ultracold atomic gases by Raman
coupling of a pair of atomic hyperfine ground states. This opens up
possibilities of simulating exotic quantum matter featuring magnetic
and spin\textendash orbit effects for ultracold atoms. Despite an
unprecedented controllability of ultracold atoms, the experimentally
realized SOC \cite{Lin2011,Zhang2012,Wang2012,Cheuk2012,Williams2012,LeBlanc2013,Engels2013,Fu2014}
couples the atomic motion to its spin just in a single spatial direction.
Such a one-dimensional (1D) SOC corresponds to an equally-weighted
combination of the Rashba- and Dresselhaus-type of coupling \cite{Liu2009,Lin2011,Li2012,Li2013}.

Realization of the synthetic SOC in two or more dimensions is highly
desirable. The two dimensional spin-orbit coupling of the Rashba type
has a non-trivial dispersion. It contains a Dirac cone at an intersection
point of two dispersion branches, as well as a highly degenerate ground
state (the Rashba ring), the latter leading to an unusual Bose-Einstein
condensation \cite{Wang2010,Xu2011,Liu2012PRA,Radic2011,Zhou2011,Sinha2011,Ozawa2012a,Ozawa2012b,Liu2013PRA,Han2015,Su2015}.
Recently, a number of elaborate schemes has been suggested to create
an effective two- and three-dimensional (2D and 3D) SOC \cite{Stanescu2007,Jacob2007,Stanescu2008,Juzeliunas2008PRA,Chuanwei-Zhang10,Campbell2011,Su2012,Anderson2012PRL,Xu2013,Anderson2013,Goldman2014,Sun2015PRA}.
In particular, Campbell \textit{et al}. proposed a way to generate
the Rashba-type SOC by cyclically coupling $N$ atomic internal states
via the Raman transitions leading to a closed-loop (ring coupling)
scheme \cite{Campbell2011}.

A variant of such a scheme has been very recently experimentally implemented
\cite{Huang2016,Meng2015} using a far detuned tripod setup corresponding
to $N=3$ in the ring coupling scheme \footnote{Two dimensional spin-orbit coupling has also been recently realized
using another approach which relies on optical lattices, see Z.~Wu,
L.~Zhang, W.~Sun, X.-T.~Xu, B.-Z.~Wang, S.-C.~Ji, Y.~Deng, S.~Chen,
X.-J.~Liu and J.-W.~Pan, arXiv:1511.08170.}. A Dirac cone \cite{Huang2016} and its opening \cite{Meng2015}
have been observed in the dispersion. However it does not seem realistic
to observe the ground-state phases associated with the Rashba ring
using the far detuning tripod setup which involves short-lived higher
hyperfine ground states \cite{Huang2016,Meng2015}. Furthermore the
$N=3$ scheme used in the experiments \cite{Huang2016,Meng2015} converges
slower to the Rashba ring than the $N=4$ ring coupling scheme \cite{Campbell2011}.

Recently Sun \textit{et al} \cite{Sun2015PRA} put forward a scheme
for generating a 2D SOC in a bilayer two-component BEC subjected to
the Raman transitions and laser-assisted interlayer tunneling. In
such a geometry the layer index provides an auxiliary degree of freedom
to form a basis of four spin-layer states. It is noteworthy that the
2D SOC provided by such a bilayer setup does not represent the Rashba
SOC \cite{Sun2015PRA}. Furthermore in such a setup the Raman transitions
should be accompanied by a recoil in different directions for different
layers, whereas the interlayer laser-assisted tunneling is to be
accompanied by a recoil in different directions for different spin
states \cite{Sun2015PRA}. These requirements would be extremely difficult
to implement experimentally.

Here we consider an alternative bilayer scheme which is free from
the above mentioned drawbacks and can provide a 2D SOC of the Rashba
type. An essential element of the bilayer scheme is that the atoms
now pick up a $\pi$ phase after completing a cyclic transition between
the four spin-layer states. In that case the dressed states of the
atom-light coupling are two-fold degenerate. As a result, one effectively
implements the $N=4$ ring coupling scheme \cite{Campbell2011} by
using a combination of two layers and two internal atomic states.
If the interlayer tunneling and Raman transitions are sufficiently
strong, the laser recoil induces a 2D SOC of the Rashba type for a
pair of degenerate atomic dressed states. In that case the minimum
of the single particle dispersion represents a degenerate Rashba-ring.

A characteristic feature of the bilayer system is that the interaction
takes place between atoms belonging to the same layer. Therefore the
atom-atom interaction is now different from the one featured for the
scheme involving four cyclically coupled atomic internal states \cite{Campbell2011}.
It is demonstrated that the bilayer scheme provides a diverse ground-state
phase diagram. In particular, in the regime of a strong atom-light
coupling the stripe and plane-wave phases emerge at specific directions
of the degenerate Rashba-ring. Thus the system exhibits an interaction-induced
anisotropy. On the other hand, in an intermediate range of the atom-light
coupling, two interlacing lattices of half-skyrmions and half-antiskyrmions
are formed.

\begin{figure}
\includegraphics[width=1\columnwidth]{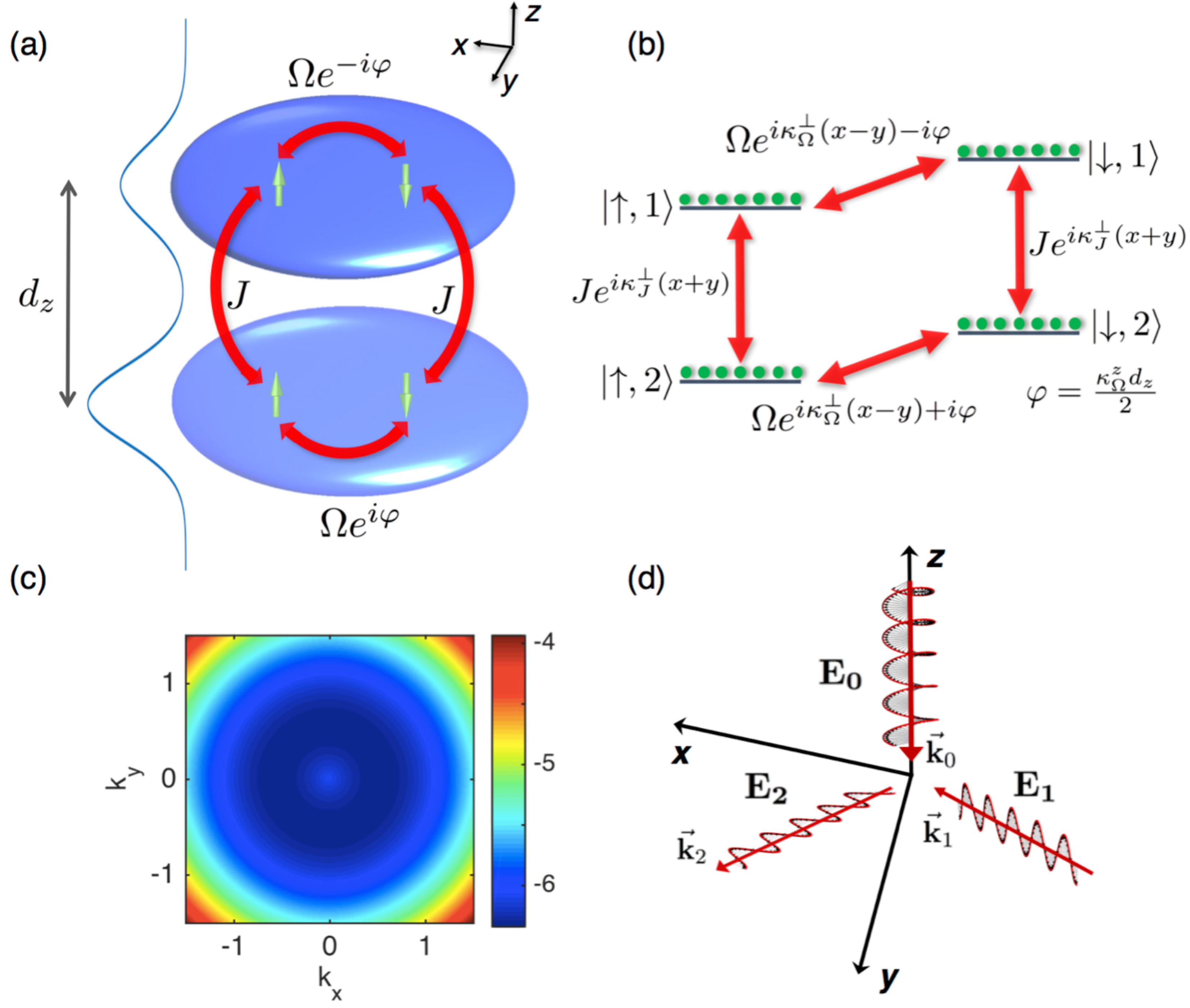}\protect\caption{(Color online) (a) Schematic plot of the atomic system. The BEC is
tightly trapped in an asymmetric double-well potential along the $z$ axis,
forming a bilayer structure. The bosonic atoms in each layer are condensed
into two single-particle internal states $|\gamma\rangle=|\uparrow\rangle,\,|\downarrow\rangle$.
The layer index $j=1,2$ provides an extra degree of freedom, so the
four states $|\gamma,j\rangle$ serve as the required atomic states
in the $N=4$ ring-coupling scheme \cite{Campbell2011}. The intralayer
transitions, $|\uparrow,j\rangle\leftrightarrow|\downarrow,j\rangle$,
are engendered by Raman coupling, while the interlayer transitions,
$|\uparrow,1\rangle\leftrightarrow|\uparrow,2\rangle$ and $|\downarrow,1\rangle\leftrightarrow|\downarrow,2\rangle$
are due to the laser-assisted tunneling. (b) Schematic plot of the
intralayer Raman transition and interlayer laser-assisted tunneling.
(c) The lowest branch of the single-particle spectrum Eq.~(\ref{eq:ground-state-dis})
for a strong symmetric coupling $\Omega=J=5E_{\mathrm{rec}}$ and
$\varphi=\pi/2$. The spectrum is plotted in units of recoil momentum
$\kappa$ and recoil energy $E_{\mathrm{rec}}$. In this case, a nearly
degenerate Rashba-ring minimum with a radius $\kappa/2$ emerges.
(d) A possible way to induce Raman transitions and interlayer tunneling
by illuminating both layers with three laser beams, two of them $\mathbf{E}_{1}$
and $\mathbf{E}_{2}$ propagating in the $xy$ plane, the third one
$\mathbf{E}_{0}$ being along the $z$ axis. The frequencies of the
laser beams are chosen such that the $\mathbf{E}_{0}$ and $\mathbf{E}_{1}$
drive the Raman transition whereas $\mathbf{E}_{0}$ and $\mathbf{E}_{2}$
induces the laser-assisted interlayer tunneling. The field $\mathbf{E}_{0}$
provides the $z$ component to the Raman coupling needed to have the
phase difference $2\varphi$ for the Raman coupling in different layers.
For more details see Appendix~A. }

\label{Fig1} 
\end{figure}

The proposed bilayer setup can be experimentally implemented by using
the current experimental technology. Unlike in the previous bilayer
scheme \cite{Sun2015PRA}, now the Raman coupling in each layer is
accompanied by a recoil in the same direction $\mathbf{e}_{x}-\mathbf{e}_{y}$
in the $xy$ plane, as one can see in Fig.~\ref{Fig1}. Consequently
each layer is affected by the Raman coupling used previously to produce
a 1D SOC \cite{Lin2011}. The $\pi$ phase shift can be realized if
the Raman coupling has an out of plane momentum component $k_{\Omega}^{z}$,
such that the relative phase between the layers is $k_{\Omega}^{z}d_{z}=2\varphi=\pi$,
where $d_{z}$ is an interlayer separation. On the other hand, the
interlayer tunneling is accompanied by the recoil in the same direction
$\mathbf{e}_{x}+\mathbf{e}_{y}$ in the $xy$ plane for both spin
states. Such an laser-assisted interlayer tunneling is also experimentally
available \cite{Aidelsburger2013,Miyake2013}. To implement the present
bilayer setup one needs to combine the Raman coupling between the
different spin states \cite{Lin2011} together with the laser-assisted
interlayer tunneling \cite{Aidelsburger2013,Miyake2013}. An additional
merit of the bilayer scheme is that only two atomic spin states are
involved. Thus there is no need to make use of spin states belonging
to a higher hyperfine manifold \cite{Campbell2011}. The latter spin
states suffer from a collisional population decay \cite{Hung2015}
undermining the effective SOC.

The paper is organized as follows. In Sec.~\ref{II}, we construct
the single-particle Hamiltonian describing spin-orbit-coupling in
a bilayer BEC affected by the atom-light interaction. The single particle
energy spectrum and corresponding eigenstates are determined for an
arbitrary strength of atom-light coupling. In Sec.~\ref{III}, we
consider the many-body ground-state phases of weakly interacting bilayer
BECs by numerically solving the Gross-Pitaevskii equations in a wide
range of magnitudes of the interatomic interaction and the atom-light
coupling. In the limit of strong atom-light coupling, we also
analyze a behavior of the ground-state phase using a variational approach,
and find it in a good agreement with the numerical results. In Sec.~\ref{IV}
we present the concluding remarks and discuss possibilities of the
experimental implementation of the proposed bilayer scheme. Finally
some auxiliary calculations are placed in Appendixes A and B.

\section{Bilayer BEC affected by the atom-light interaction\label{II}}

\subsection{Single-particle Hamiltonian}

To realize the synthetic SOC in the atomic BEC based on the $N=4$
close-loop (ring-coupling) scheme \cite{Campbell2011}, we consider
a two-component Bose gas confined in the bilayer geometry depicted
in Fig.~\ref{Fig1}. The atoms are confined in a deep enough
asymmetric double-well potential \cite{Sebby-Strabley2006}, so their
motion is suppressed in the $z$-direction. The atoms are
in the ground states of individual wells, and only the laser-assisted
tunneling can induce transitions between the two wells. The four combined
spin-layer states $\left|\gamma,j\right\rangle \equiv|\gamma\rangle_{\mathrm{spin}}\otimes|j\rangle_{\mathrm{layer}}$
serve as the states required for the ring coupling scheme \cite{Campbell2011}.
Here $j=1,\,2$ signifies the $j$-th layer, and $|\gamma\rangle=|\uparrow\rangle,\,|\downarrow\rangle$
denotes an internal (quasi-spin) atomic state. The spin-layer states
are cylindrically coupled by illuminating the atoms by three lasers
inducing the intralayer Raman transitions and the laser-assisted interlayer
tunneling, as depicted in Fig.~\ref{Fig1}. As it is shown in
Appendix~\ref{A}, the resultant single-particle Hamiltonian
can be represented as 
\begin{equation}
\hat{H}_{0}=\hat{H}_{\mathrm{atom}}+\hat{H}_{\mathrm{intra}}+\hat{H}_{\mathrm{inter}}+\hat{H}_{\mathrm{extra}}\,,\label{H_0}
\end{equation}
where 
\begin{equation}
\hat{H}_{\mathrm{atom}}=\int d^{2}\mathbf{r}_{_{\bot}}\sum_{j,\gamma}\hat{\psi}_{\gamma j}^{\dag}\frac{\hbar^{2}\mathbf{k}_{_{\bot}}^{2}}{2m}\hat{\psi}_{\gamma j}\,,\label{H00}
\end{equation}
is a Hamiltonian for an unperturbed atomic motion within the layers,
\begin{equation}
\hat{H}_{\mathrm{intra}}=\int d^{2}\mathbf{\mathbf{r}_{_{\bot}}}\Omega\left[e^{i\varphi}\hat{\psi}_{\uparrow1}^{\dag}\hat{\psi}_{\downarrow1}+e^{-i\varphi}\hat{\psi}_{\uparrow2}^{\dag}\hat{\psi}_{\downarrow2}+{\rm H.c.}\right]\label{eq:H2-1}
\end{equation}
describes the spin-flip intralayer Raman transitions characterized
by the Rabi frequency $\Omega$, and 
\begin{equation}
\hat{H}_{\mathrm{inter}}=\int d^{2}\mathbf{r}_{_{\bot}}\sum_{\gamma}J\hat{\psi}_{\gamma2}^{\dag}\hat{\psi}_{\gamma1}+{\rm H.c.}\label{H3-1}
\end{equation}
represents the laser-assisted interlayer tunneling with the strength
$J$. Finally, the last term 
\begin{align}
\hat{H}_{\mathrm{extra}}= & \int d^{2}\mathbf{r}_{_{\bot}}\frac{\hbar^{2}\kappa}{m}\left[\hat{\psi}_{\uparrow2}^{\dag}k_{x}\hat{\psi}_{\uparrow2}-\hat{\psi}_{\downarrow1}^{\dag}k_{x}\hat{\psi}_{1\downarrow}\right.\nonumber \\
 & +\left.\hat{\psi}_{\downarrow2}^{\dag}k_{y}\hat{\psi}_{\downarrow2}-\hat{\psi}_{\uparrow1}^{\dag}k_{y}\hat{\psi}_{\uparrow1}\right]\,,\label{eq:H-SOC}
\end{align}
describes the spin-orbit coupling due to the recoil momentum $\kappa$
in the $xy$ plane induced by the interlayer tunneling and Raman transitions.
Here $\hat{\psi}_{\gamma j}$ is an operator annihilating an atom
with a spin $\gamma$ in the $j$th layer, $\mathbf{\mathbf{r}_{_{\bot}}}=\left(x,y\right)$
and $\mathbf{\mathbf{k}_{_{\bot}}}=\left(k_{x},k_{y}\right)$ are
in-plane projections of the atomic position vector and momentum, and
$2\varphi=k_{\Omega}^{z}d_{z}$ is a phase difference between
the Raman couplings in the two layers. The latter phase difference
can be tuned by either varying the double-well separation $d_{z}$
or the out-of-plane Raman recoil $k_{\Omega}^{z}$. To implement an
$N=4$ ring coupling scheme with a $\pi$ phase shift \cite{Campbell2011}
the Raman coupling in different layers should have a $\pi$ phase
difference, so we set $\varphi=\pi/2$ throughout the paper.

\begin{figure}
\includegraphics[width=1\columnwidth]{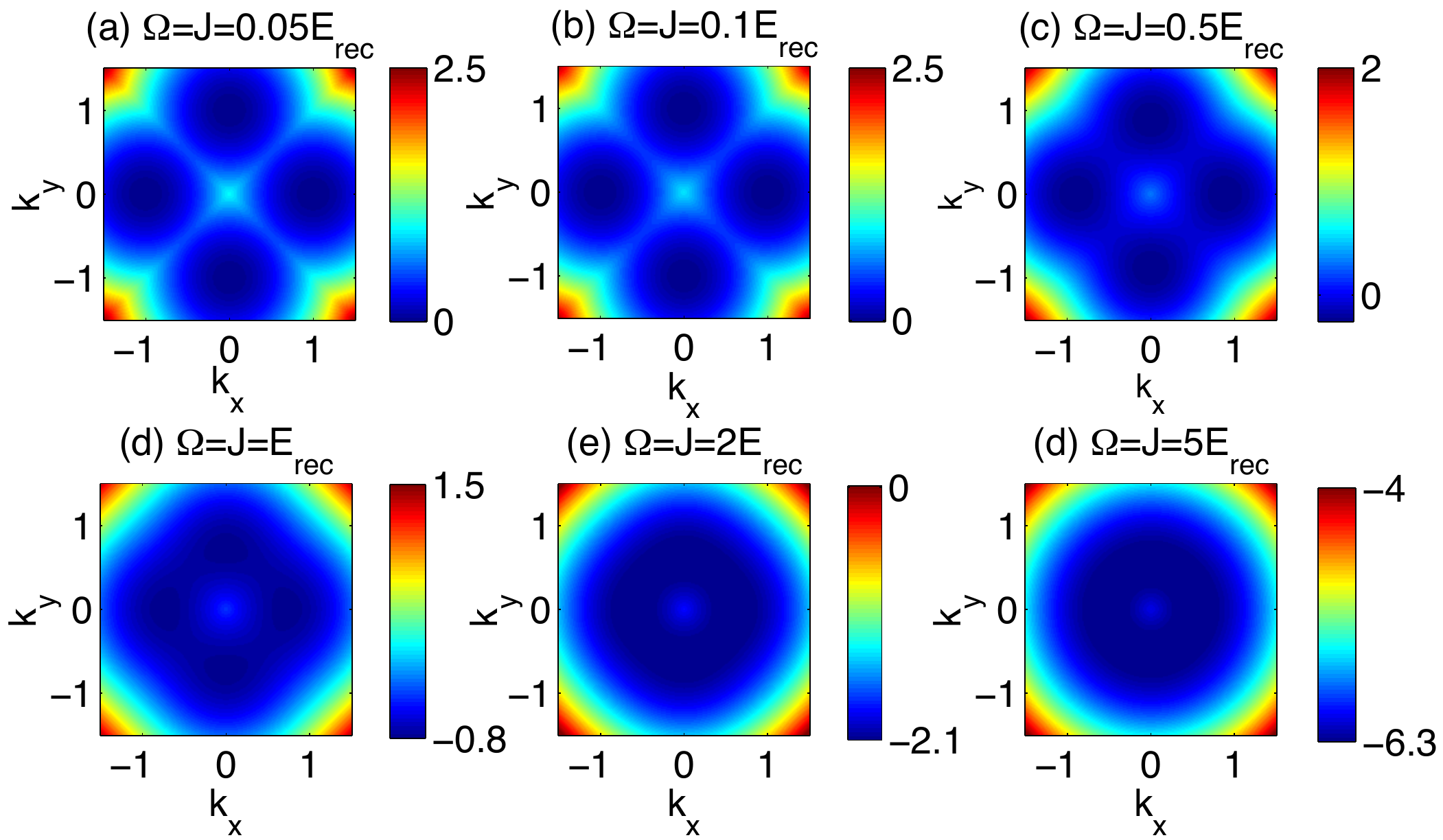}\protect\caption{(Color online) The lowest branch of the single-particle dispersion
as a function of momentum for various coupling strengths are shown
in (a)--(f). In a weak coupling regime $\Omega^2=J^2\ll E_{\mathrm{rec}}^2$,
the dispersion is a superimposition of four distinct paraboloids centered
at $\pm\kappa\hat{\mathbf{e}}_{x}$ and $\pm\kappa\hat{\mathbf{e}}_{y}$
as depicted in (a). Increasing the coupling strength, the four paraboloids
become mixed with each other as plotted in (b) and (c) for $\Omega^2=J^2\lesssim E_{\mathrm{rec}}^2$
and the minima become much shallower as shown in (d) and (e) when $\Omega^2=J^2\sim E_{\mathrm{rec}}^2$.
In the strong coupling regime $\Omega^2=J^2\gg E_{\mathrm{rec}}^2$ (see Ref.~\cite{Campbell2011}), the
Rashba-ring minimum with a radius $\kappa/2$ emerges, as one can see in (e) and (f).}

\label{Fig2} 
\end{figure}

Note that in the original representation the laser-induced terms $\hat{H}_{{\rm intra}}^{\prime}$
and $\hat{H}_{{\rm inter}}^{\prime}$ contain position dependent recoil
factors featured in Eqs.~(\ref{eq:H2}) and (\ref{eq:H3}) in 
Appendix~\ref{A}. Such a position-dependence can be eliminated via
the transformation (\ref{eq:transformation-1}) leading to a position-independent
single-particle Hamiltonian $\hat{H}_{0}$ given by Eq.~(\ref{H_0}).
Additionally, the spin-orbit coupling term $\hat{H}_{{\rm extra}}$
appears in the transformed Hamiltonian $\hat{H}_{0}$.

In the following, we shall work in dimensionless units where
the energy is measured in units of the recoil energy $E_{\mathrm{rec}}=\hbar^{2}\kappa^{2}/2m$
and the wave-vector is measured in the units of $\kappa$. 

\begin{figure}
\includegraphics[width=0.9\columnwidth]{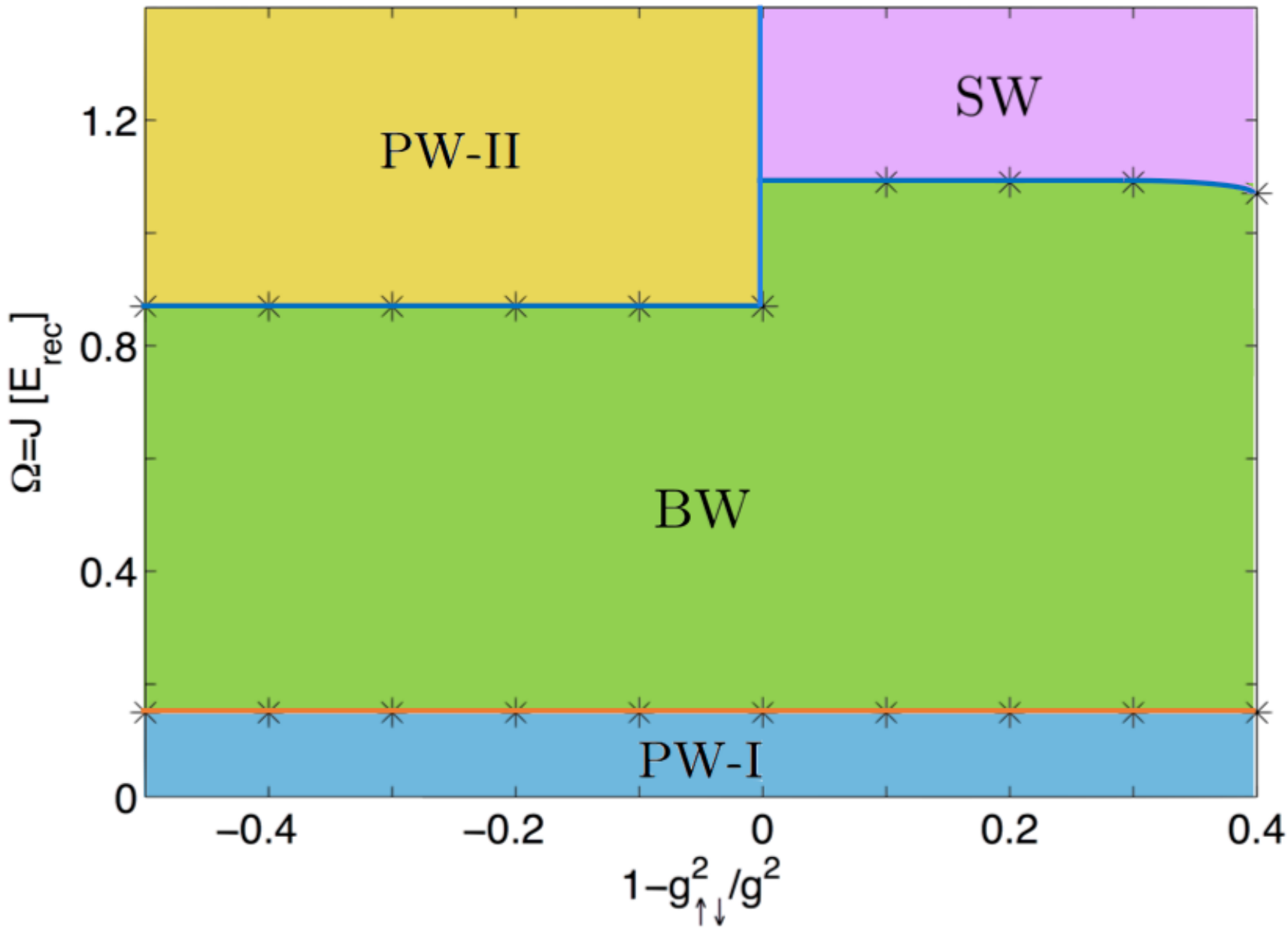}\protect\caption{(Color online) Ground-state phase diagram of the bilayer spin-orbit-coupled
BEC as a function of $1-g_{\uparrow\downarrow}^{2}/g^{2}$ and the
laser-assisted coupling $\Omega$ for $J=\Omega$. The phase diagram
consists of two types of plane-wave phases (PW-I: cyan and PW-II:
yellow), a brick-wall phase (BW: green) and a standing-wave phase
(SW: purple).}

\label{Fig3} 
\end{figure}

\begin{figure*}
\includegraphics[width=1.8\columnwidth]{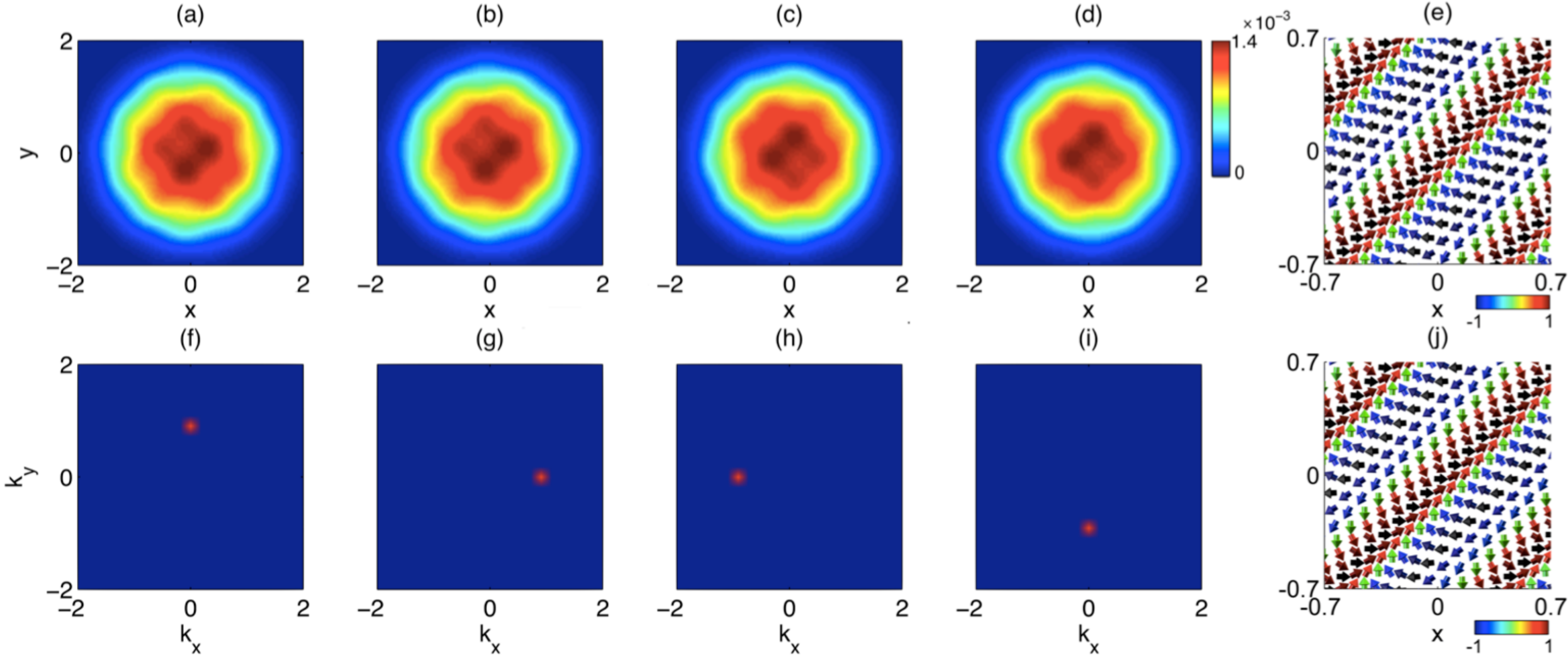}\protect\caption{(Color online) (a)--(d) Plots of the real-space density
profiles of all spin-layer components for the PW-I phase, i.e., $\rho_{\uparrow1}$,
$\rho_{\downarrow1}$, $\rho_{\uparrow2}$, and $\rho_{\downarrow2}$,
respectively. The corresponding momentum-space distributions are depicted
in (f)--(i), where the axes are in units of recoil momentum. The spin
texture in the first and second layers are shown in (e) and (j), respectively,
where the color of the arrows indicate the magnitude of $S_{j}^{x}$.
The couplings and interaction strengths are taken to be $\Omega=J=0.05E_{\mathrm{rec}}$
and $g_{\uparrow}:g_{\downarrow}:g_{\uparrow\downarrow}=1:1:0.9$.}

\label{Fig4} 
\end{figure*}

\subsection{Single-particle dispersion}

Diagonalization of the single-particle Hamiltonian [Eq.~(\ref{H_0})]
yields four branches of the single-particle dispersion considered
in Appendix B.1. Here, we focus only on the lowest branch characterized
by the eigenenergies  
\begin{equation}
E_{g}=1+k^{2}-\sqrt{\Omega^{2}+J^{2}+2k^{2}+2a_{\mathbf{k}}}\,,\label{eq:ground-state-dis}
\end{equation}
with 
\begin{equation}
a_{\mathbf{k}}=\sqrt{\Omega^{2}(k_{x}+k_{y})^{2}+J^{2}(k_{x}-k_{y})^{2}+(k_{x}^{2}-k_{y}^{2})^{2}}\,,
\end{equation}
where $\mathbf{\mathbf{k}}\equiv\mathbf{\mathbf{k}_{_{\bot}}}=\left(k_{x},k_{y}\right)$
is an atomic momentum. 

For a symmetric coupling ($\Omega=J$), the
ground-state dispersion surface is plotted in Fig.~\ref{Fig2} for
various coupling strengths. In the following we shall present the
corresponding eigenstates in different regimes of the coupling strength
at the local minima of the dispersion surface where the atoms condense.

In the weak coupling regime, $\Omega^2=J^2\ll E_{\mathrm{rec}}^2$, the
dispersion surface is built of superimposed paraboloids centered at
$\pm\kappa\mathbf{e}_{x}$ and $\pm\kappa\mathbf{e}_{y}$, as shown
in Fig.~\ref{Fig2} (a). Each eigenstate corresponding to the
four energy minima contains a single spin-layer component 
\begin{equation}
|\downarrow,1\rangle e^{i\kappa x}\,,\quad|\uparrow,2\rangle e^{-i\kappa x}\,,\quad|\uparrow,1\rangle e^{i\kappa y}\,,\quad|\downarrow,2\rangle e^{-i\kappa y}\,,\label{eq:eigenvector-weak}
\end{equation}
where  
\begin{eqnarray}
|\uparrow,1\rangle & = & \left(\begin{array}{c}
1\\
0\\
0\\
0
\end{array}\right)\,,\quad|\downarrow,1\rangle=\left(\begin{array}{c}
0\\
1\\
0\\
0
\end{array}\right)\,,\label{eq:columns-main-text}\\
|\uparrow,2\rangle & = & \left(\begin{array}{c}
0\\
0\\
1\\
0
\end{array}\right)\,,\quad|\downarrow,2\rangle=\left(\begin{array}{c}
0\\
0\\
0\\
1
\end{array}\right)\,,\nonumber 
\end{eqnarray}
represents a basis of the spin-layer states. Therefore the four
spin components are not yet mixed in the weak coupling limit.

With increasing the coupling to $\Omega^2=J^2\lesssim E_{\mathrm{rec}}^2$,
the four paraboloids gradually coalesce but still the dispersion exhibits
four distinguishable minima located at $\pm\kappa\mathbf{e}_{x}$
and $\pm\kappa\mathbf{e}_{y}$ as depicted in Figs.~\ref{Fig2}(b)--\ref{Fig2}(d).
Each eigenstate corresponding to the four energy minima now contains
contributions of three spin states 
\begin{equation}
\left(\begin{array}{c}
1\\
\frac{2i}{\Omega}\\
0\\
-i
\end{array}\right)e^{i\kappa x}\,,\,\left(\begin{array}{c}
1\\
0\\
\frac{-2}{J}\\
i
\end{array}\right)e^{-i\kappa x}\,,\,\left(\begin{array}{c}
\frac{-2i}{\Omega}\\
1\\
i\\
0
\end{array}\right)e^{i\kappa y}\,,\,\left(\begin{array}{c}
0\\
1\\
-i\\
\frac{-2}{J}
\end{array}\right)e^{-i\kappa y}\,.\label{eq:eigenvector-moderate}
\end{equation}
This will lead to a brickwall phase for the bilayer BEC.

Finally in the strong coupling regime, $\Omega^2=J^2\gg E_{\mathrm{rec}}^2$,
one has $a_{\mathbf{k}}\approx\Omega k\sqrt{2}$ and thus
$E_{g}\approx-\Omega\sqrt{2}+1-k+k^{2}$. Hence mixing between the spin states results
in the emergence of a cylindrically symmetric Rashba-ring minimum
of a radius $\kappa/2$ in the dispersion shown in Fig.~\ref{Fig2}
(e) and (f). This is a characteristic feature of the close-loop (ring
coupling) scheme \cite{Campbell2011}. In this regime, the 
single particle eigenstates $\Psi_{\mathbf{k}_{g}}$ on the Rashba-ring
takes the form 
\begin{equation}
\chi=\left(\begin{array}{c}
\sqrt{2}\cos\phi\\
i(1-\sin\phi+\cos\phi)\\
1-\sin\phi-\cos\phi\\
-\sqrt{2}i(1-\sin\phi)
\end{array}\right)\frac{e^{i\mathbf{k}\cdot\mathbf{r}_{_{\bot}}}}{\sqrt{8-8\sin\phi}}\,,\label{eq:eigenvector-ring}
\end{equation}
with $\mathbf{k}=\mathbf{k}_{g}=\kappa(\cos\phi\,\mathbf{e}_{x}+\sin\phi\,\mathbf{e}_{y})/2$,
where $\phi$ is an azimuthal angle parameterizing the degenerate
ring. 

It is convenient to project the system onto the state-vectors $\chi^{(1)}$
and $\chi^{(2)}$ corresponding to the spinor part of the ground-state-vector
(\ref{eq:eigenvector-ring}) for $\phi=3\pi/4$ and $\phi=-\pi/4$,
i.e. corresponding to the opposite momenta $\mathbf{k}$ and $\mathbf{-k}$ along the diagonal $\mathbf{e}_{x} - \mathbf{e}_{y}$,
see Eq.(\ref{eq:rotated-basis-low-energy}) in Appendix B.2. The projected
Hamiltonian represents a Rashba-type Hamiltonian given by Eq.(\ref{eq:H-Rashba}).

Note that the Rashba-ring minimum occurs only for a symmetric
coupling where $\Omega=J$. The asymmetric coupling $(\Omega\ne J)$
breaks the rotational symmetry in the momentum space, reducing the
ring minimum to a two-fold degenerate ground-state.

\section{Mean-Field Ground States\label{III}}

\subsection{Gross-Pitaevskii energy functional}

We assume that all atoms interact with each other via contact potentials.
As a result, the second-quantized interaction Hamiltonian is given by
\begin{equation}
\hat{H}_{\mathrm{int}}=\int d^{2}\mathbf{r}_{_{\bot}}\sum_{j=1,2}\left(\frac{g_{\uparrow}}{2}\hat{n}_{\uparrow j}^{2}+\frac{g_{\downarrow}}{2}\hat{n}_{\downarrow j}^{2}+g_{\uparrow\downarrow}\hat{n}_{\uparrow j}\hat{n}_{\downarrow j}\right)\,,\label{intH}
\end{equation}
where the interlayer interaction is neglected because of the short-range
nature of the interatomic interactions. Here $g_{\uparrow}$ and $g_{\downarrow}$
denote the intraspecies interaction strengths, $g_{\uparrow\downarrow}$
is the interspecies interaction strength, and $\hat{n}_{\gamma j}=\hat{\psi}_{\gamma j}^{\dag}\hat{\psi}_{\gamma j}$
is the number density operator for the $\gamma$-th spin state in the
$j$-th layer. 
To approach
the ground-state structure of the spin-orbit coupled BEC at zero temperature,
we adopt the mean-field approximation, namely, the field operator,
$\hat{\psi}_{\gamma j}$, is replaced by the ground-state expectation
value, $\psi_{\gamma j}\equiv\langle\hat{\psi}_{\gamma j}\rangle$,
which is complex in general. Accordingly, the Gross-Pitaevskii (GP)
energy functional $\mathcal{E}[\psi_{\gamma j}^{*},\psi_{\gamma j}]=\langle\hat{H}_{0}+\hat{H}_{{\rm \mathrm{int}}}\rangle$
is explicitly expressed as 
\begin{align}
\mathcal{E}[\psi_{\gamma j}^{*},\psi_{\gamma j}]= & \int d^{2}\mathbf{r}_{_{\bot}}\biggl[\sum_{j,\gamma}\psi_{\gamma j}^{*}\left(-\frac{1}{2}\nabla_{\perp}^{2}+\frac{1}{2}\omega^{2}r^{2}\right)\psi_{\gamma j}\nonumber \\
 & +\kappa(\psi_{\uparrow2}^{*}\hat{p}_{x}\psi_{\uparrow2}-\psi_{\downarrow1}^{*}\hat{p}_{x}\psi_{\downarrow1})\nonumber \\
 & +\kappa(\psi_{\downarrow2}^{*}\hat{p}_{y}\psi_{\downarrow2}-\psi_{\uparrow1}^{*}\hat{p}_{y}\psi_{\uparrow1})\nonumber \\
 & +\Omega\left(e^{i\varphi}\psi_{\uparrow1}^{*}\psi_{\downarrow1}+e^{-i\varphi}\psi_{\uparrow2}^{*}\psi_{\downarrow2}+{\rm H.c.}\right)\nonumber \\
 & +J(\psi_{\uparrow2}^{*}\psi_{\uparrow1}+\psi_{\downarrow2}^{*}\psi_{\downarrow1}+{\rm H.c.})\nonumber \\
 & +\sum_{j}\left(\frac{g_{\uparrow}}{2}\rho_{\uparrow j}^{2}+\frac{g_{\downarrow}}{2}\rho_{\downarrow j}^{2}+g_{\uparrow\downarrow}\rho_{\uparrow j}\rho_{\downarrow j}\right)\biggr]\,,\label{G-P}
\end{align}
where $\rho_{\gamma j}=|\psi_{\gamma j}|^{2}$, and the wave functions
are normalized to the unity $\int d^{2}\mathbf{r}_{_{\bot}}\sum_{j\gamma}\rho_{\gamma j}(\mathbf{r}_{_{\bot}})=1$.
This is achieved by the substitution $\psi_{\gamma j}\rightarrow\sqrt{N}\psi_{\gamma j}$ which
rescales interaction strengths, viz., $g_{\uparrow\downarrow}\rightarrow Ng_{\uparrow\downarrow}$,
$g_{\uparrow}\rightarrow Ng_{\uparrow}$ and $g_{\downarrow}\rightarrow Ng_{\downarrow}$,
where $N$ is the total number of atoms. Without loss of generality, we assume $g_{\uparrow}=g_{\downarrow}\equiv g$. Furthermore to confine atoms we have included
a sufficiently weak harmonic trapping potential with a energy $\hbar\omega$
much smaller than the recoil energy $E_{\mathrm{rec}}$.

An important quantity characterizing the bilayer BEC is a spin texture on the $j$-th layer $\mathbf{S}_{j}(\mathbf{r}_{_{\bot}})=\langle\chi_{j}|\boldsymbol{\sigma}|\chi_{j}\rangle$ \cite{Kasamatsu2005},
where $\boldsymbol{\sigma}=\sigma_x \mathbf{e}_x+\sigma_y \mathbf{e}_y+\sigma_z \mathbf{e}_z$ is a vector of Pauli matrices, and
$\chi_{j}(\mathbf{r}_{_{\bot}})=[\chi_{\uparrow j}(\mathbf{r}_{_{\bot}}),\chi_{\downarrow j}(\mathbf{r}_{_{\bot}})]^{T}$ 
is a  local spinor.
The latter $\chi_{j}(\mathbf{r}_{_{\bot}})$ is proportional to the spinor wave-function $\psi_{\gamma j}(\mathbf{r}_{_{\bot}})=\sqrt{\sum_{\gamma}\rho_{\gamma j}(\mathbf{r}_{_{\bot}})}\chi_{\gamma j}(\mathbf{r}_{_{\bot}})$
and is normalized to unity $|\chi_{\uparrow j}|^{2}+|\chi_{\downarrow j}|^{2}=1$.
It is convenient to represent the spinor $\chi_{j}(\mathbf{r}_{_{\bot}})$  in terms of its amplitude and phase 
\begin{equation}
\chi_{\gamma j}(\mathbf{r}_{_{\bot}})=|\chi_{\gamma j}|e^{i\theta_{\gamma j}}, \quad \mathrm{with} \quad \gamma=\uparrow, \downarrow\,.
\label{eq:chi}
\end{equation}
 In that case the Cartesian components of the vector $\mathbf{S}_{j}$
take the form
\begin{eqnarray}
S_{j}^{x} & = & 2|\chi_{\uparrow j}||\chi_{\downarrow j}|
\cos(\theta_{\downarrow j}-\theta_{\uparrow j})\,,\nonumber \\
S_{j}^{y} & = & 2|\chi_{\uparrow j}||\chi_{\downarrow j}|
\sin(\theta_{\downarrow j}-\theta_{\uparrow j})\,,\nonumber \\
S_{j}^{z} & = & |\chi_{\uparrow j}|^{2}-|\chi_{\downarrow j}|^{2}\,.\label{eq:spin-texture}
\end{eqnarray}

\begin{figure}
\includegraphics[width=1\columnwidth]{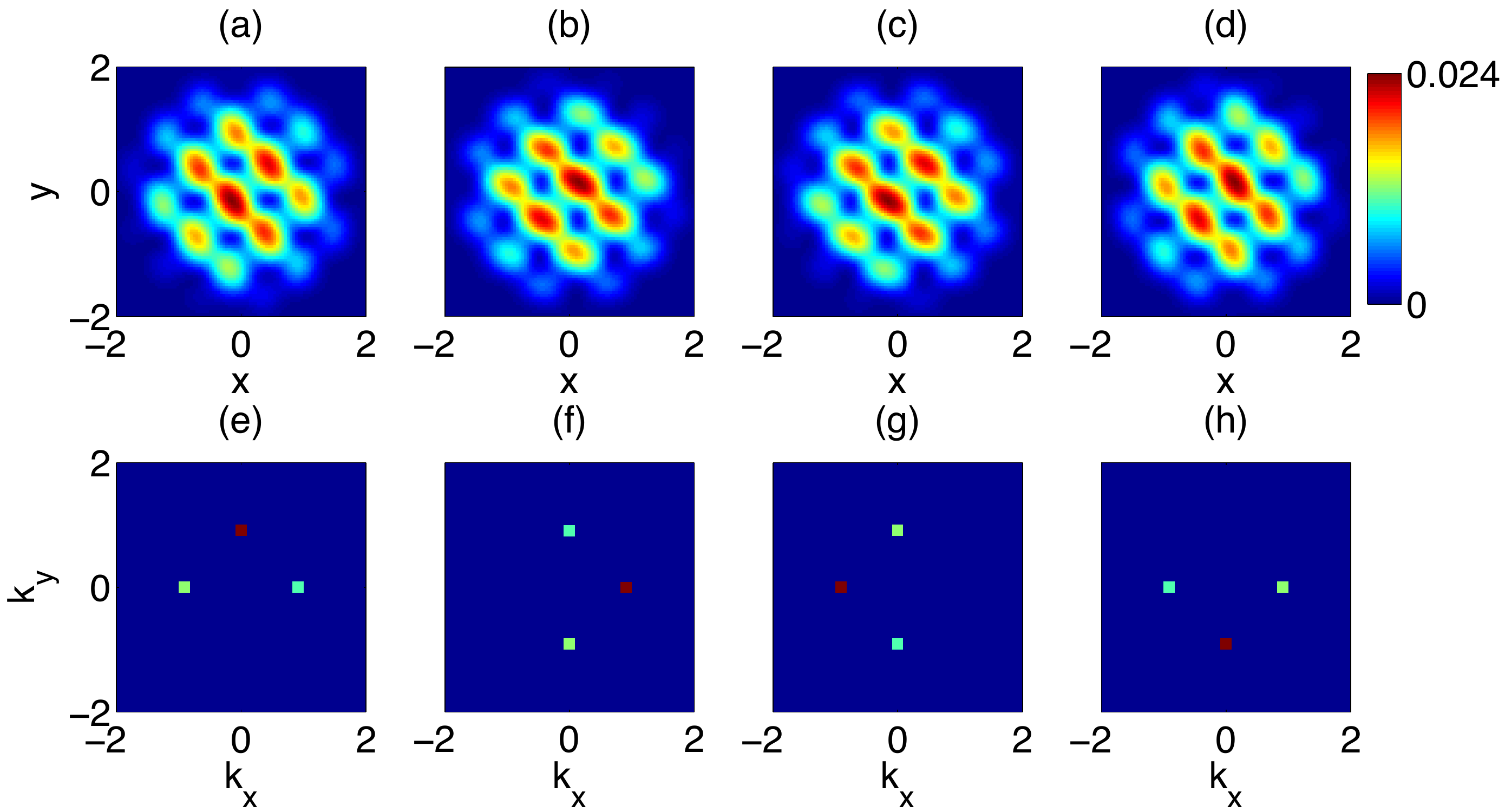}\protect\caption{(Color online) The real-space density profiles of all spin-layer components
in BW phase, $\rho_{\uparrow1}$, $\rho_{\downarrow1}$, $\rho_{\uparrow2}$
and $\rho_{\downarrow2}$, are plotted in (a)-(d), respectively. The
corresponding momentum-space distributions are depicted in (e)-(h),
where the axes are calibrated in units of recoil momentum. The couplings
and interaction strengths are taken to be $\Omega=J=0.5E_{\mathrm{rec}}$
and $g_{\uparrow}:g_{\downarrow}:g_{\uparrow\downarrow}=1:1:0.9$.}

\label{Fig5} 
\end{figure}

\subsection{Numerical results}

To investigate the ground-state phases of the interacting BEC in a
harmonic trap, we minimize the GP energy functional Eq.~(\ref{G-P})
by the imaginary-time propagation method \cite{Chin2005}. As shown in Fig.~\ref{Fig3},
the ground state possesses a variety of phases which are determined
by the inter- and intralayer coupling and the intralayer interaction between the atoms.
In the numerical simulations, four distinct phases have been identified. 
These are the plane-wave phases of types I and II (PW-I and PW-II), the
brick-wall (BW) phase, as well as the standing-wave (SW) or stripe phase.
The occurrence of PW-I and BW phases depends only on the Raman coupling and
the interlayer tunneling. On the other hand, the PW-II and SW phases emerge at stronger
Raman coupling and stronger tunneling, and depend on the interatomic interactions.
In the following, the structure of each phases is discussed.

\begin{figure}
\includegraphics[width=0.95\columnwidth]{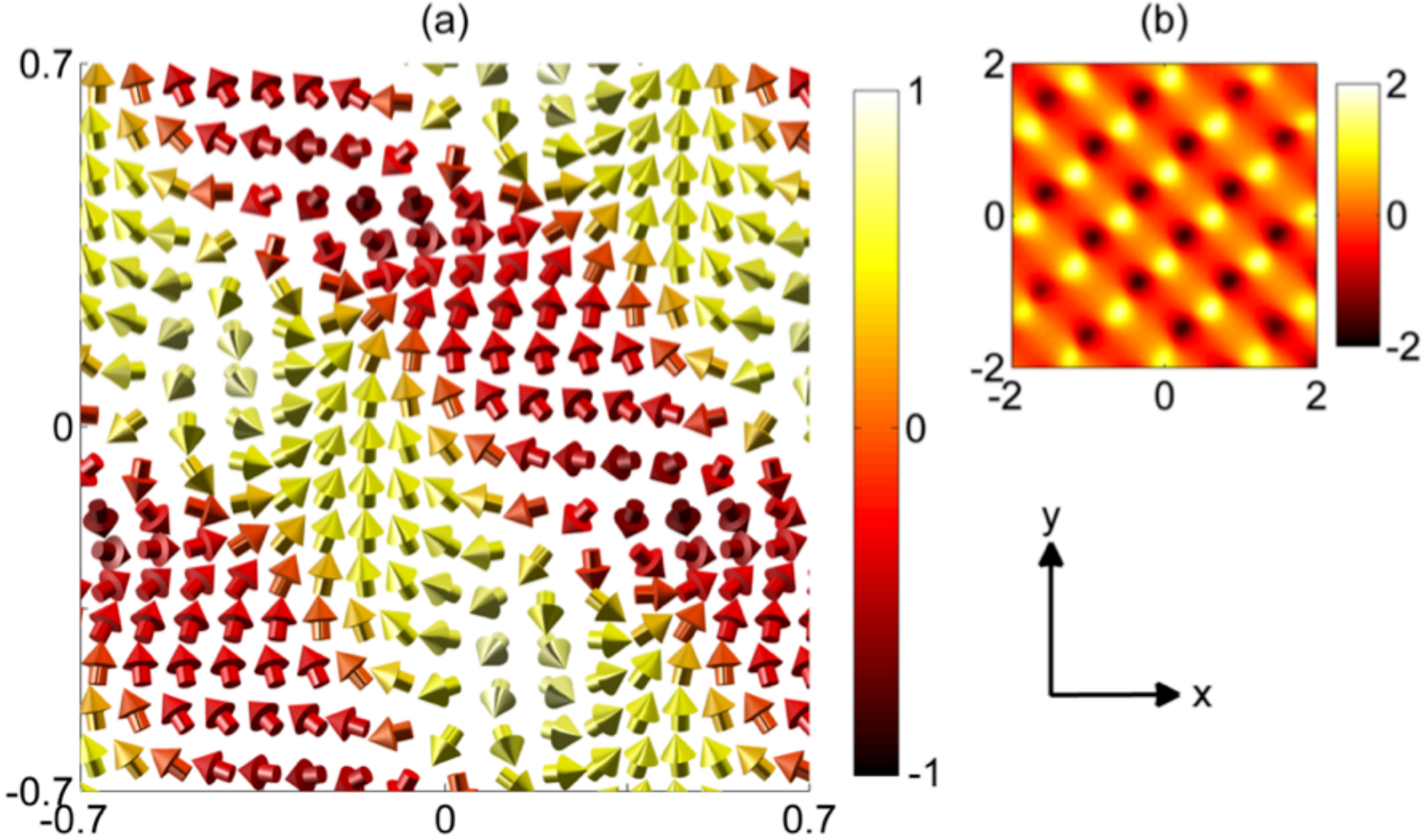}\protect\caption{(Color online) (a) Spin texture of the first layer for the BW phase
depicted in Fig.~\ref{Fig5}. The color of the arrows indicates the
magnitude of $S_{1}^{z}$. (b) The topological charge density of the
spin orientation shown in (a). Two interlacing square lattices of
positive and negative charges are clearly visible. Integrating the
charge density over an unit cell for the lattice of positive (negative)
charge gives $1/2$ ($-1/2$) which corresponds to the half-skyrmion
(half-antiskyrmion).}

\label{Fig6} 
\end{figure}

\textit{PW-I phase.} In a weak coupling regime, $\Omega^2=J^2\ll E_{\mathrm{rec}}^2$,
the four spin-layer components are almost uncoupled. Consequently each layer
behaves like an ordinary binary BEC except that the single-particle
dispersion is shifted due to the term $\hat{H}_{\mathrm{extra}}$, Eq.~(\ref{eq:H-SOC}), induced by the gauge-transformation (\ref{eq:transformation-1}).
Therefore, each spin-layer component would condense at the bottom
of the shifted parabolic dispersion, as shown in Fig.~\ref{Fig4}.
The real-space density profiles of the four spin-layer components,
$|\psi_{\gamma j}(\mathbf{r}_{_{\bot}})|^{2}$, are presented in Fig.~\ref{Fig4}
(a)--(d), and their momentum-space counterparts, $|\bar{\psi}{}_{\gamma j}(\mathbf{k})|^{2}$
, are shown in Fig.~\ref{Fig4} (f)--(i). The momentum distribution of each component, $|\bar{\psi}{}_{\gamma j}(\mathbf{k})|^{2}$,
is sharply peaked around the four momenta, $\mathbf{k}=\kappa\mathbf{e}_{x}$,
$-\kappa\mathbf{e}_{x}$, $\kappa\mathbf{e}_{y}$ and $-\kappa\mathbf{e}_{y}$, indicating that each spin-layer component acquires a
momentum shift via the SOC term $\hat{H}_{\mathrm{extra}}$ given by Eq.~(\ref{eq:H-SOC}).

The spin texture $\mathbf{S}_{j}(\mathbf{r}_{_{\bot}})$ of PW-I phase is depicted in Figs.~\ref{Fig4}(e)
and \ref{Fig4}(j) for the first layer ($j=1$). The color of the arrows indicates the magnitude of $S_{j}^{x}$
and the periodic modulation of spin orientation is caused by the interference
between the plane waves characterizing the spin-layer components.

\begin{figure}
\includegraphics[width=0.85\columnwidth]{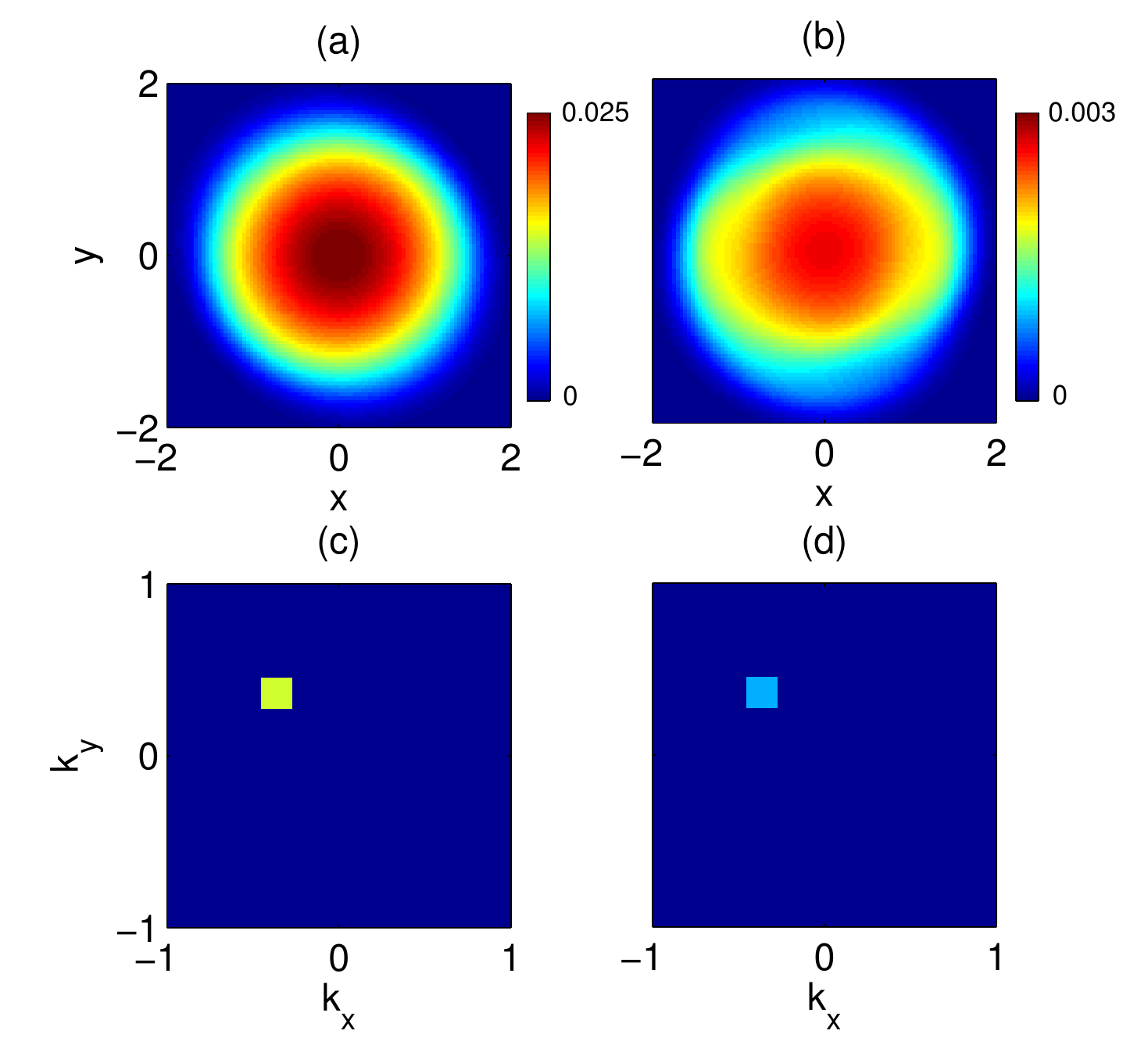}\protect\caption{(Color online) (a) and (b) The real-space density profiles of the spin-layer components
$\rho_{\uparrow1}$ and $\rho_{\downarrow1}$
in the first layer for PW-II phase with  $\mathbf{k}=\kappa\mathbf{e}_{-}/2$. The corresponding momentum-space
distributions are depicted in (c) and (d), where the axes are marked in
units of the recoil momentum. The couplings and interaction strengths
are taken to be $\Omega=J=2E_{\mathrm{rec}}$ and $g_{\uparrow}:g_{\downarrow}:g_{\uparrow\downarrow}=1:1:1.1$.}

\label{Fig7} 
\end{figure}
\begin{figure*}
\includegraphics[width=1.8\columnwidth]{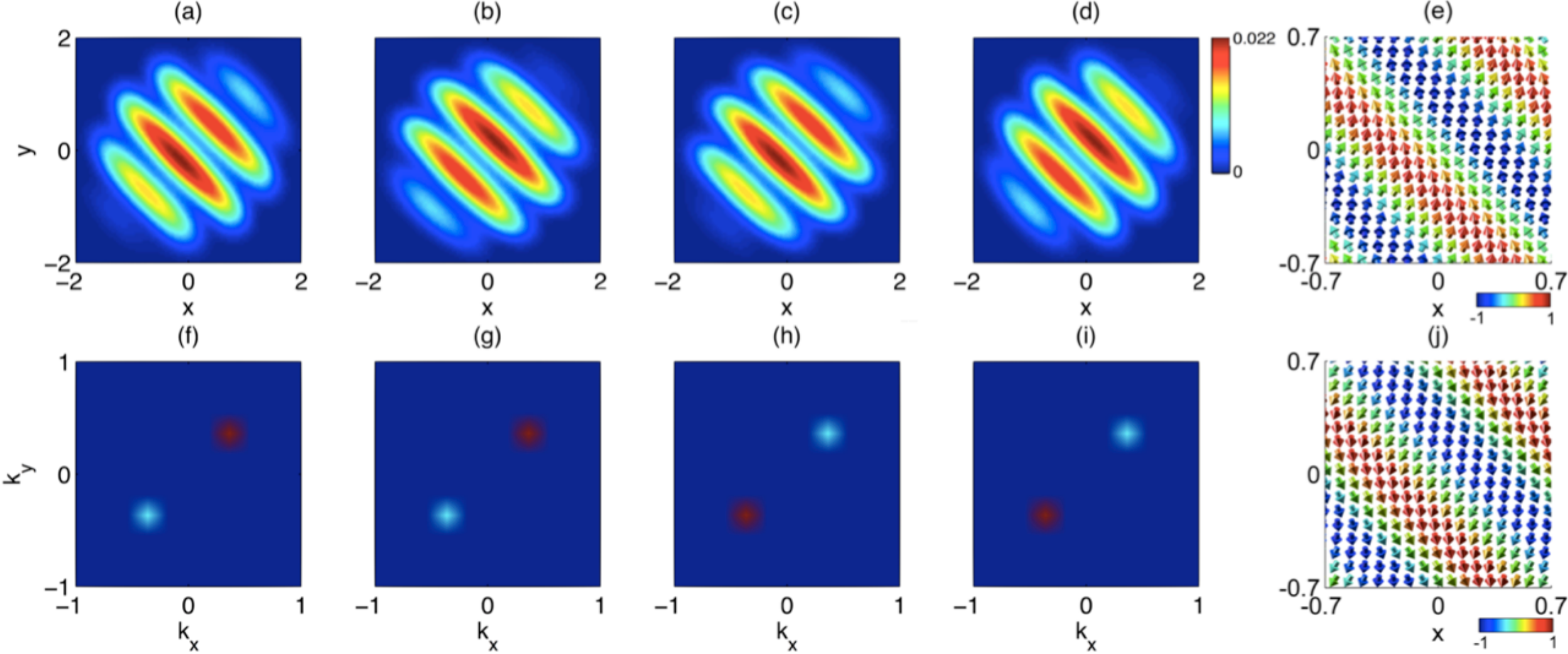}\protect\caption{(Color online) The real-space density profiles of all spin-layer components
in SW phase, $\rho_{\uparrow1}$, $\rho_{\downarrow1}$, $\rho_{\uparrow2}$
and $\rho_{\downarrow2}$, are plotted in (a)--(d), respectively.
The corresponding momentum-space distributions are depicted in (f)--(i),
where the axes are calibrated in units of recoil momentum. The spin
texture in the first and second layers are shown in (e) and (j), respectively
where the color of the arrows indicate the magnitude of $S_{j}^{z}$.
The couplings and interaction strengths are taken to be $\Omega=J=2E_{\mathrm{rec}}$
and $g_{\uparrow}:g_{\downarrow}:g_{\uparrow\downarrow}=1:1:0.9$.}
\label{Fig8} 
\end{figure*}
\textit{BW phase.} By simultaneously increasing $\Omega$ and $J$,
the four otherwise distinct paraboloids characterizing the PW-I phase
start developing a noticeable overlap between the neighboring paraboloids
and finally completely merge in the moderate coupling regime where
$\Omega^2=J^2\lesssim E_{\mathrm{rec}}^2$. The dispersion surface so formed
introduces a ground-state phase shown in Figs.~\ref{Fig5}(a)-\ref{Fig5}(d).
The BEC density profiles of the four spin-layer components now exhibit
periodic spatial modulations characteristic to a BW pattern. Note
that the dips in the density profiles are not vortices according to
their phase profiles. The BW patterns of both spin components in the
same layer interlace, so that the density dips of one spin component are filled by 
another spin component. The formation of BW structure can
be easily understood by examining the density profiles in the momentum
space. 

As shown in Figs.~\ref{Fig5}(e)-\ref{Fig5}(h), it is evident that $\bar{\psi}{}_{\gamma j}(\mathbf{k})$
appears as a superposition of three out the four-momentum eigenmodes
labeled by $\mathbf{k}=\pm\kappa\mathbf{e}_{x}$ and $\pm\kappa\mathbf{e}_{y}$.
For instance, let us take $\psi_{\uparrow1}$ representing a superposition of
the modes with $\mathbf{k}=\pm\kappa\mathbf{e}_{x}$ and $\kappa\mathbf{e}_{y}$.
In this case the majority of atoms condense in the $\mathbf{k}=\kappa\mathbf{e}_{y}$
mode, whereas the remaining atoms evenly condense in the $\mathbf{k}=\pm\kappa\mathbf{e}_{x}$
modes. The latter two modes are populated owing to the presence of
non-negligible inter- and intralayer couplings. 

In contrast to the PW-I phase, the BW structure leads to an intriguing
spin texture in each layer, as shown in Fig.~\ref{Fig6}. The spin
texture consists of two interlacing square lattices of spin vortices
with opposite handednesss. To further characterize this state, we
calculate the topological charge density in the $j$-th layer, $\tau_{j}=\mathbf{S}_{j}\cdot\partial_{x}\mathbf{S}_{j}\times\partial_{y}\mathbf{S}_{j}/4\pi$.
As shown in Fig.~\ref{Fig6}, the left (right)-handed circulation
corresponds to a positive (negative) topological charge density. Integrating
$\tau_{j}$ over the elementary unit cell, we identify that the topological
charge can be either $+1/2$ or $-1/2$. This corresponds to the half-skyrmions
and half-antiskyrmions, respectively \cite{Mermin1976,Mizushima2002,Su2012,Sun2015PRA}.

\textit{PW-II phase.} Now let us assume that $g_{\uparrow\downarrow}>g$ and consider the strong coupling limit
where $\Omega^2=J^2\gg E_{\mathrm{rec}}^2$. In this regime, the Rashba-ring
minimum emerges, and the many-body ground state (PW-II phase) becomes interaction-dependent.
Figure~\ref{Fig7} illustrates formation of the PW-II phase for $\Omega=J=2E_{\mathrm{rec}}$
and $g_{\uparrow}:g_{\downarrow}:g_{\uparrow\downarrow}=1:1:1.1$ corresponding to the case where $g_{\uparrow\downarrow}>g$.
Unlike in the PW-I phase, here
each spin-layer component condenses in the same momentum mode with $\mathbf{k}=\pm\kappa\mathbf{e}_{-}/2$ along the diagonal $\mathbf{e}_{-}=(\mathbf{e}_{x}-\mathbf{e}_{y})/\sqrt{2}$,
and the total multicomponent wave function contains a common plane-wave factor.
In each layer the intralayer spin polarization is non-zero due to
the imbalanced population of $\rho_{\uparrow j}$ and $\rho_{\downarrow j}$.
On the other hand, the density profiles of the same spin-component
but different layers are identical, i.e., $\rho_{\gamma1}=\rho_{\gamma2}$.
Note that in the previously considered Rashba-type SO-coupled system \cite{Wang2010},
the plane wave phase exists in the regime where $g_{\uparrow\downarrow}<g$.
This is opposite to the current bilayer system.

\textit{SW phase.}-- Finally, for $g_{\uparrow\downarrow}<g$ and
$\Omega^2=J^2\gg E_{\mathrm{rec}}^2$, the ground-state wave function consists
of two counterpropagating plane-waves on the Rashba ring
with opposite momenta along the diagonal $\mathbf{e}_{+}=(\mathbf{e}_{x}+\mathbf{e}_{y})/\sqrt{2}$. 
This constitutes the 
SW phase. As shown in Fig.~\ref{Fig8}, the real-space density profile
of each component with $\Omega=J=2E_{\mathrm{rec}}$ and $g_{\uparrow}:g_{\downarrow}:g_{\uparrow\downarrow}=1:1:0.9$
forms the stripe structure, while the momentum-space density is sharply
peaked around the of two momenta $\mathbf{k}=\pm\kappa\mathbf{e}_{+}/2$.
The spin texture in each layer is depicted in Figs.~\ref{Fig8}(e)
and \ref{Fig8}(j). The periodic modulation of the spin texture is accompanied
by the stripe structure of the density profile. Furthermore, one can see in Fig.~\ref{Fig8} that the occupation
of the two momentum states $\mathbf{k}=(\pm\kappa/2)\mathbf{e}_{+}$ is asymmetric in the bilayer system,
in contrast to the SW phase in the previously considered SO-coupled BECs 
\cite{Wang2010,Sinha2011,Ozawa2012a,Ozawa2012b,Chen2014PRA,Su2015}.
It is noteworthy that now the SW phase occurs for $g_{\uparrow\downarrow}<g$.
This is opposite to the usual BEC affected by the Rashba SOC \cite{Wang2010}. 
To further understand the the phases of the bilayer system, a variational analysis
is presented in the following section.

\subsection{Variational approach}

So far our conclusions on the BEC phases were based mostly based on numerical simulations.
In order to gain a better insight into the ground-state structure of the
bilayer SO-coupled BEC, a simpler analytical study is desirable. To this
end, we employ a variational approach to investigate the ground-state
phases in different coupling regimes. We are  particularly interested
in solving the ground-state in the strong-coupling regime, where the
many-body ground state shows a preference of residing at some special
locations  of the degenerate Rashba ring.

We begin by writing down the interaction energy, namely, the ground-state expectation
value of the  interaction Hamiltonian (\ref{eq:E_int}) 
\begin{eqnarray}
\mathcal{E}_{\mathrm{int}} & = & \frac{1}{4}\sum_{j}\left(c_{0}\rho_{j}^{2}+c_{2}\mu_{j}^{2}\right),\label{eq:E_int}
\end{eqnarray}
where $\rho_{j}=\rho_{\uparrow j}+\rho_{\downarrow j}$ and $\mu_{j}=\rho_{\uparrow j}-\rho_{\downarrow j}$
are respectively the total number and magnetization densities in the
$j$-th layer, and $c_{0}=g+g_{\uparrow\downarrow}$ and $c_{2}=g-g_{\uparrow\downarrow}$
characterize the density-density and spin-spin interactions,
respectively. We use the following
trial wave functions of PW-II and SW phases 
\begin{eqnarray}
\Psi^{\mathrm{PW-II}} & = & \Psi_{\mathbf{k}_{g}}\label{eq:PW-II}
\end{eqnarray}
and 
\begin{eqnarray}
\Psi^{\mathrm{SW}} & = & \frac{1}{\sqrt{2}}(\Psi_{\mathbf{k}_{g}}+\Psi_{\mathbf{-k}_{g}}),\label{eq:SP}
\end{eqnarray}
where $\Psi_{\mathbf{k}_{g}}$ is the plane-wave solution given by Eq.~(\ref{eq:eigenvector-ring}), with $\mathbf{k}_{g}=\kappa(\cos\phi\,\mathbf{e}_{x}+\sin\phi\,\mathbf{e}_{y})/2$.
In the following, we compare the interaction energies for these
two trial wave functions. For simplicity, we shall not include a harmonic trapping potential.

\textit{PW-II phase.} Let us first consider the variational ansatz of PW-II
phase. With the trial wave function given by Eq.~(\ref{eq:PW-II}), the total
density and spin density in the $j$-th layer read 
\begin{equation}
\rho_{j}=\frac{1}{2}+\frac{(-1)^{j-1}(\cos\phi+\sin\phi)}{4}
\end{equation}
and 
\begin{equation}
\mu_{j}=\frac{\sin\phi-\cos\phi}{4}\,.\label{eq:spin_densities_variation}
\end{equation}
Therefore, for the PW-II phase the nonlinear interaction energy is given by 
\begin{equation}
\mathcal{E}_{\mathrm{int}}^{\mathrm{PW-II}}=\frac{c_{0}}{8}+\frac{c_{0}}{64}(1+\sin2\phi)+\frac{c_{2}}{32}(1-\sin2\phi)\,.\label{eq:E_int_PW_II}
\end{equation}
It is evident that the interaction energy depends on the azimuthal angle $\phi$. This is in contrast to the single-layer Rashba SO-coupled system in which the interaction energy does not depend on the azimuthal angle $\phi$ 
\cite{Wang2010,Sinha2011,Chen2014PRA}. Using Eq.~(\ref{eq:E_int_PW_II}), the energy minima are found at two angles $\phi=3\pi/4$ and $-\pi/4$, for which
\begin{eqnarray}
\mathcal{E}_{\mathrm{int},\,\mathrm{min}}^{\mathrm{PW-II}} & = & \frac{c_{0}}{8}+\frac{c_{2}}{16}\,.\label{eq:E_int_PW_II_min}
\end{eqnarray}

\textit{SW phase.} For the SW phase, 
the trial wave function Eq.~(\ref{eq:SP}) provides the following total density and magnetization density: 
\begin{equation}
\rho_{j}=\frac{1}{2}+\frac{1}{4}|\cos\phi|(\tan\phi-1)\cos(x\cos\phi+y\sin\phi)\label{eq:density_variation_SW}
\end{equation}
and 
\begin{equation}
\mu_{j}=\frac{1}{4}|\cos\phi|(1+\tan\phi)\cos(x\cos\phi+y\sin\phi),\label{eq:spin_density_variation_SW}
\end{equation}
where the spatial dependence comes from the periodic modulation of
the stripes. The resultant energy takes the form 
\begin{eqnarray}
\mathcal{E}_{\mathrm{int}}^{\mathrm{SW}} & = & \frac{c_{0}}{8}+\frac{c_{0}}{64}(1-\sin2\phi)+\frac{c_{2}}{64}(1+\sin2\phi),\label{eq:E_int_SP}
\end{eqnarray}
where the spatially oscillating cosine terms are replaced by the mean
values $\langle\cos(x\cos\phi+y\sin\phi)\rangle=0$ and $\langle\cos^{2}(x\cos\phi+y\sin\phi)\rangle=1/2$.
Thus the interaction is again anisotropic along the Rashba ring.
The energy minimum occurs at $\phi=\pi/4$ or equivalently at $5\pi/4$ 
\begin{eqnarray}
\mathcal{E}_{\mathrm{int},\,\mathrm{min}}^{\mathrm{SW}} & = & \frac{c_{0}}{8}+\frac{c_{2}}{32}.\label{eq:E_int_SP_min}
\end{eqnarray}

The energy minima of the PW-II and SW phases differ by the amount $\mathcal{E}_{\mathrm{int},\,\mathrm{min}}^{\mathrm{PW-II}}-\mathcal{E}_{\mathrm{int},\,\mathrm{min}}^{\mathrm{SW}}=c_{2}/32$. This implies that for $c_{2}>0$ ($c_{2}<0$) the SW (PW-II) phase represents
the ground state, in agreement with the numerical simulations. Although the ordinary 
single layer Rashba SOC
also provides the SW and PW-II phases \cite{Wang2010}, the conclusions are
opposite compared to our bilayer system, that is, for $c_{2}<0$ ($c_{2}>0$) the ground
state is in the SW (PW-II) phase. 

We note that in a single-layer Rashba SOC the energy of the PW phase is spin-independent on a Rashba ring, and the phase has a zero magnetization~\cite{Wang2010}. On the other hand, in the bilayer system the energy minima of PW-II  phase on the Rashba ring are characterized by a non-vanishing magnetization. Therefore the PW-II phase has a lower energy than the SW phase for  $c_{2}<0$  corresponding to $g_{\uparrow\downarrow}>g$. In this way one arrives at a situation opposite to that appearing for an ordinary single layer BEC affected by the SOC [15] in which the PW phase has an energy lower than the SW phase if $c_{2}>0$ corresponding to $g_{\uparrow\downarrow}>g$.  The difference originates from the anisotropy in the population of each spin-layer component on the Rashba ring in the bilayer system, as one can see in~Eq.~(\ref{eq:eigenvector-ring}).

In this way, the variational approach shows that the atoms favor to condense at
 $\phi=3\pi/4$ or $-\pi/4$ for the PW-II phase, whereas the SW phase
 involves a superposition of the plane waves at  $\phi=\pi/4$ and $5\pi/4$. To gain more insight into such an interaction-induced
symmetry breaking, in the Appendix~B.3 the Hamiltonian has been expressed in terms of the basis vectors 
of the lowest dispersion branch at the azimuthal angles $\phi=3\pi/4$ or $-\pi/4$. The projection of the Hamiltonian onto these states
gives rise to the appearance of the Rashba Hamiltonian (\ref{eq:H-Rashba}) subjected to an 
asymmetric atom-atom interaction given by Eq.(\ref{eq:e-2}).

We have presented the variational study in the regime of strong coupling.
For a weak and a moderate coupling, the single-particle dispersion
surfaces are characterized by four distinct minima in the momentum
space. The trial wave functions are then simply superpositions of the
four corresponding momentum eigenstates. This provides the ground state phases
in a good agreement with the numerical results.

\subsubsection*{Asymmetric coupling $\Omega\protect\ne J$}

Now let us briefly discuss a situation when $\Omega\ne J$ and
$\sqrt{\Omega^{2}+J^{2}}\gg E_{\mathrm{rec}}$. The asymmetric coupling
breaks the rotational symmetry in the momentum space and leads to a two-fold
degenerate single-particle dispersion. For $\Omega>J$, the energy
minimum occurs at $\phi=\pi/4$ and $5\pi/4$, while for $\Omega<J$,
the energy minimum appears at $\phi=-\pi/4$ and $3\pi/4$. Numerical
simulations reveal that the many-body ground state is the SW or PW-II
phase for $\Omega>J$ or $\Omega<J$, respectively. The phase is independent
of the interaction strengths. This is because the single particle dispersion
possesses only two degenerate minima. In other words, due to the lack of the ring degeneracy,
the many-body ground state is determined only by the strengths of the
Raman coupling and the tunneling.

\subsubsection*{Metastable states}

\begin{figure}
\includegraphics[width=0.9\columnwidth]{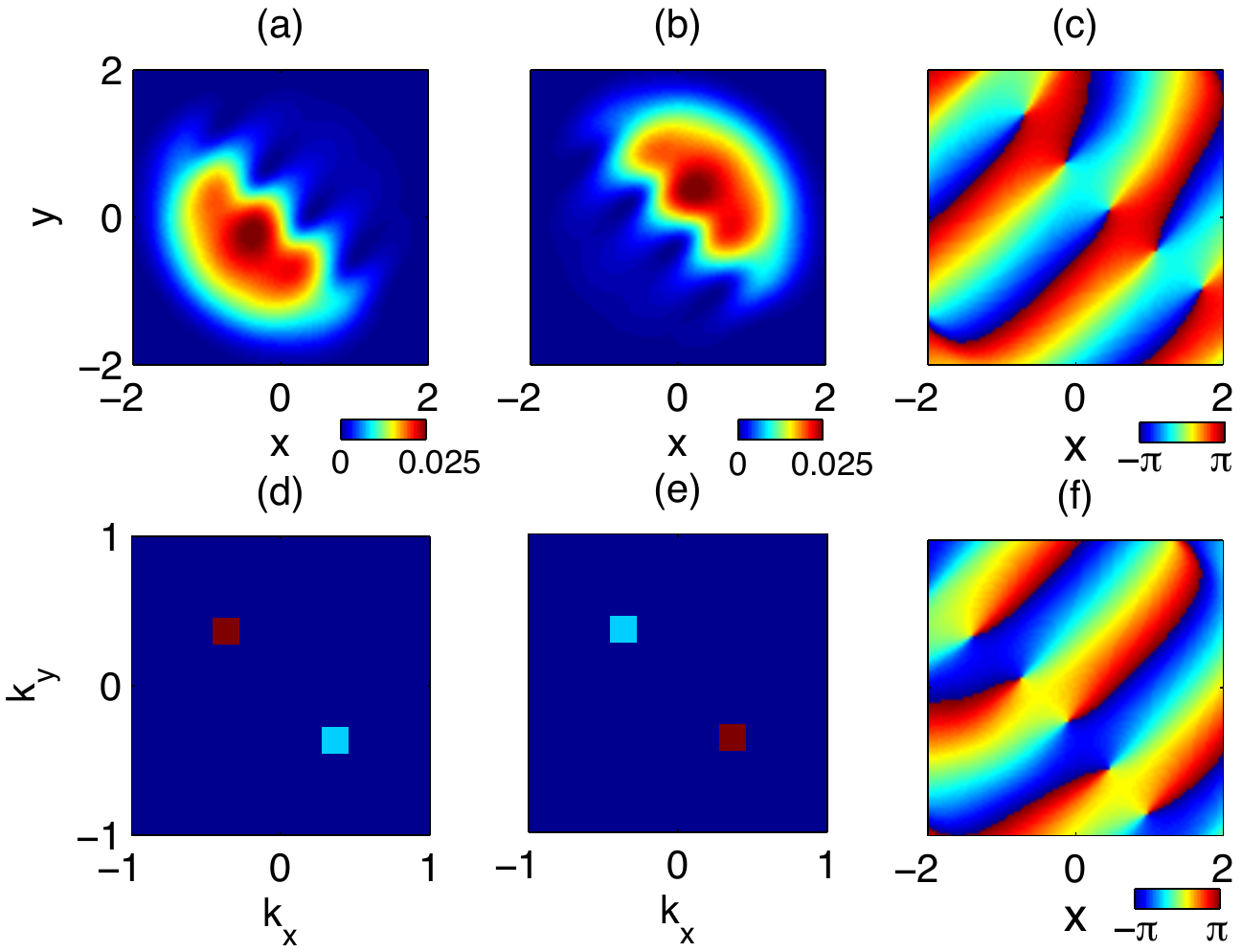}\protect\caption{(Color online) The real-space density profiles of spin components
in the first layer in a metastable state, $\rho_{\uparrow1}$ are $\rho_{\downarrow1}$,
are plotted in (a)--(b), respectively. The corresponding momentum-space
distributions are depicted in (d)--(e), where the axes are calibrated
in units of recoil momentum. The coupling is now asymmetric, $\Omega=1.5E_{rec}$ and $J=2E_{rec}$,  and interaction strengths
are  $g_{\uparrow}:g_{\downarrow}:g_{\uparrow\downarrow}=1:1:0.9$.
The phase profiles of the wave function in the first layer, $\theta_{\uparrow1}(\mathbf{r})$
and $\theta_{\downarrow1}(\mathbf{r})$ are plotted in (c) and (f),
respectively, where an array of vortices can be clearly seen.}

\label{Fig9} 
\end{figure}

Occasionally the imaginary-time propagation 
ended up at a metastable state containing domains. The metastable state emerges for parameters of the system corresponding to the PW-II phase, namely,
 for  $c_{2}<0$ with $\Omega^2=J^2\gg E_{\mathrm{rec}}^2$ or for $\Omega\ne J$
with $\sqrt{\Omega^{2}+J^{2}}\gg E_{\mathrm{rec}}$. It has an energy
slightly higher than that of the ground state. The metastable state
is made of two spatially separated spin-polarized domains in the
same layer, as shown in Fig.~\ref{Fig9}. The domains carry opposite momenta,
$\mathbf{k}=\pm\kappa\mathbf{e}_{-}/2$, like in the previously considered
case of the ordinary Rashba SOC \cite{Wang2010}. 
Since the
phases of the two counterpropagating PW-II states in each domain could
not continuously connect along the boundary, the frustration results
in the formation of arrays of vortices, as depicted in Figs.~\ref{Fig9}(c) and \ref{Fig9}(f). The density of the vortices increases with increasing
the SOC strength.

\section{Discussion and Conclusion}

\label{IV}

In conclusion, the proposed bilayer system provides a possibility
to realize the Rashba-type SOC for ultracold atoms. Numerical simulation 
and variational analysis have elucidated
a diverse phase diagram of the bilayer BEC  in a wide range
of magnitudes of the atom-light coupling and atom-atom interaction.
In the moderate coupling regime the BW phase is formed leading to 
the emergence of lattices of half-skyrmions and half-antiskyrmions. 
In the strong coupling
regime, the Rashba-ring minimum emerges, and the ground state is
either the SW or PW-II phases, depending on the interatomic interaction
strengths. 

An experimental implementation of the proposed bilayer SO-coupled system is within
reach of current experiments with ultracold atoms. For instance,
the two magnetic sub-levels of the $F=1$ ground state manifold of
the $^{87}\mathrm{Rb}$-type alkali atoms \cite{Lin2011} could serve
as the atomic internal (quasi-spin 1/2) states. Typically the experimental trapping frequencies
are $(\omega_{\bot},\omega_{z})=2\pi\times\left(10,400\right)$ Hz
and the wavelength of laser fields inducing the Raman coupling and interlayer tunneling is around
$\lambda_{L}\simeq800$ nm,
corresponding to the recoil energy $E_{\mathrm{rec}}\simeq11\hbar\omega_{\bot}$.
The scattering lengths for the two spin states $|F=1,m_{F}=0\rangle\equiv|\uparrow\rangle$
and $|F=1,m_{F}=-1\rangle\equiv|\downarrow\rangle$, used in Ref.~\cite{Lin2011},
are given by $a_{\uparrow}=c_{0}$ and $a_{\downarrow}=a_{\uparrow\downarrow}=c_{0}+c_{2}$,
with $c_{0}=7.79\times10^{-12}\,\mathrm{Hz}\,\mathrm{cm}^{3}$ and
$c_{2}=-3.61\times10^{-14}\,\mathrm{Hz}\,\mathrm{cm}^{3}$ \cite{Ho1998,Ohmi1998}.
The intra and interspecies interaction strengths are given by $g_{\uparrow,\downarrow}=\sqrt{2\pi}Na_{\uparrow,\downarrow}/\xi_{z}$
and $g_{\uparrow\downarrow}=\sqrt{2\pi}Na_{\uparrow\downarrow}/\xi_{z}$
with $\xi_{z}=\sqrt{\hbar/m\omega_{z}}$. The corresponding intra-species
interaction is nearly symmetric with $g_{\uparrow}/g_{\downarrow}=1.0047$,
so the phase diagram of Fig.~\ref{Fig3} can be applied directly.
Finally, the diverse phase diagram of the bilayer system also provides
the possibilities to study the quantum phase transition by varying
the coupling strengths which will be investigated in another study. 
\begin{acknowledgments}
SWS and SCG are supported by Ministry of Science and Technology, Taiwan
(Grants No. MOST 103-2112- M-018- 002-MY3). SCG is also supported
by National Center for Theoretical Sciences, Taiwan. WML is supported
by NSFC (Grants Nos. 11434015, 61227902, 61378017), by NKBRSFC (Grants
Nos. 2011CB921502, 2012CB821305), by SKLQOQOD (Grants No. KF201403),
and by SPRPCAS (Grants No. XDB01020300). GJ is supported by Lithuanian
Research Council (Grants No. MIP-086/2015). ACJ and QS are supported
by NSFC (Grants No. 11404225 and No. 11474205). LW is supported by
NSFC (Grants Nos. 11504037). 
\end{acknowledgments}

\appendix

\section{Atom-Light Interaction\label{A}}

In this Appendix, we provide a full account of the atom-light interaction
processes proposed for generating SOC in a bilayer BEC. The general
Hamiltonian $H_{{\rm AL}}$ of the atom-light interaction in an atomic
hyperfine ground-state manifold is expressed in terms of the scalar
and vector light shifts \cite{Goldman2014,Deutsch1998}: 
\begin{align}
H_{{\rm AL}}= & u_{s}(\mathbf{E}^{*}\cdot\mathbf{E})+\frac{iu_{v}g_{F}}{\hbar g_{J}}(\mathbf{E}^{*}\times\mathbf{E})\cdot\hat{\mathbf{F}}\,,\label{eq:light-potential}
\end{align}
where $\mathbf{E}^{*}$ is negative frequency part of the full electric
field, $\hat{\mathbf{F}}$ the total spin operator, and $u_{s}$ and
$u_{v}$ are the scalar and vector atomic polarizabilities. The parameters
$g_{J}$ and $g_{F}$ denote the Landé g-factors due to the electronic
spin and the total angular momentum of the atom, respectively. For
$^{87}{\rm Rb}$ atoms in the lowest energy hyperfine manifold with
$F=1$, one has $g_{F}/g_{J}=-1/4$. Additionally, the atoms are trapped
in a spin-independent asymmetric double-well potential \cite{Sebby-Strabley2006}.
The energy difference for the atomic ground states localized in different
layers is $\Delta_{\mathrm{inter}}$, whereas the Zeeman splitting
between atomic internal spin states within a layer is $\Delta_{\mathrm{intra}}$.

Figure~\ref{Fig1} (d) illustrates the laser configuration for creating
the desirable intra- and interlayer couplings. As shown in Fig.~\ref{Fig1}
(a), both layers are simultaneously illuminated by three laser beams
labeled by $\mathbf{E}_{0}$, $\mathbf{E}_{1}$, and $\mathbf{E}_{2}$.
The former field $\mathbf{E}_{0}\sim(\mathbf{e}_{x}+i\mathbf{e}_{y})e^{i(k_{0}z-\omega_{0}t)}$
is circularly polarized and propagates along the $z$-axis.
It contributes both to the intra- and inter-layers coupling.
The latter fields $\mathbf{E}_{1}$ and $\mathbf{E}_{2}$ are responsible
for producing the intra- and interlayer couplings, respectively. In
the following we shall consider these couplings in more details.

\subsection{Intralayer transitions}

The other applied field, $\mathbf{E}_{1}\sim\hat{\mathbf{e}}_{z}e^{i[\mathbf{k}_{1}\cdot\mathbf{r}-(\omega_{0}+\delta\omega_{1})t]}$,
is linearly polarized along $\hat{\mathbf{e}}_{z}$ and is characterized
by the wave vector $\mathbf{k}_{1}=k_{1}\mathbf{e}_{-}$ in the $xy$-plane,
as one can see in Fig.~\ref{Fig1} (d), where $\mathbf{e}_{-}=(\mathbf{e}_{x}-\mathbf{e}_{y})/\sqrt{2}$.
The vector product $\mathbf{E}_{0}^{*}\times\mathbf{E}_{1}$ in Eq.~(\ref{eq:light-potential})
describes the intralayer spin-flip transitions taking place if the
frequencies of the fields $\mathbf{E}_{0}$ and $\mathbf{E}_{1}$
are tuned to the two-photon resonance, $\delta\omega_{1}=\Delta_{\mathrm{intra}}$,
between the magnetic sublevels $|\downarrow\rangle\equiv|m_{F}=-1\rangle$
and $|\uparrow\rangle\equiv|m_{F}=0\rangle$. The third magnetic sublevel
$|m_{F}=1\rangle$ can be excluded due to a sufficiently large quadratic
Zeeman effect, as demonstrated by the NIST group \cite{Lin2011}.
Therefore, the Hamiltonian of the intralayer Raman coupling can be
written as 
\begin{align}
\hat{H}_{{\rm intra}}^{\prime}= & \int d^{2}\mathbf{r}_{_{\bot}}\sum_{j}[\Omega e^{i\mathbf{(\mathbf{k}_{\Omega}^{\bot}}\cdot\mathbf{r}_{_{\bot}}+(-1)^{j}\varphi-\delta\omega_{1}t)}+\mathrm{c.c.}]\nonumber \\
 & \times\hat{\Phi}_{\uparrow\textit{j}}^{\dag}\hat{\Phi}_{\downarrow\textit{j}}+\mathrm{H.c.}\,,\label{eq:H2}
\end{align}
where $\mathbf{k}_{\Omega}=\mathbf{k}_{1}-\mathbf{k}_{0}=k_{\Omega}^{\bot}\mathbf{e}_{-}+k_{\Omega}^{z}\mathbf{e}_{z}$
with $k_{\Omega}^{\perp}=k_{1}$ and $k_{\Omega}^{z}=k_{0}$. Here
$\hat{\Phi}_{\gamma j}(\mathbf{r}_{_{\bot}},z)$ is a field operator
annihilating an atom in the spin-layer state $\left|\gamma,j\right\rangle $,
and $\Omega$ is the Rabi frequency of the intralayer Raman coupling.
Since the atoms move freely only the $xy$ plane, the out-of-plane
Raman recoil provides the phase difference $2\varphi=k_{\Omega}^{z}d_{z}$
for the Raman coupling in different layers. The phase difference can
be tuned by either varying the double-well separation $d_{z}$ or
the out-of-plane Raman recoil $k_{\Omega}^{z}$. In what follows,
we take $\varphi=\pi/2$ to get the $N=4$ close-loop scheme \cite{Campbell2011}.

\subsection{Interlayer tunneling}

The third applied field, $\mathbf{E}_{2}\sim\mathbf{e}_{-}e^{i[\mathbf{k}_{2}\cdot\mathbf{r}-(\omega_{0}+\delta\omega_{2})t]}$,
propagates along $\mathbf{k}_{2}=k_{2}\mathbf{e}_{+}$
with $\mathbf{e}_{+}=(\mathbf{e}_{x}+\mathbf{e}_{y})/\sqrt{2}$ and is linearly polarized along $\mathbf{e}_{-}$
in the $xy$-plane. Since $\mathbf{E}_{0}$ and $\mathbf{E}_{2}$
are not orthogonal, their scalar product $\mathbf{E}_{0}\cdot\mathbf{E}_{2}$
featured in Eq.~(\ref{eq:light-potential}), provides a scalar light shift
oscillating with a frequency $\delta\omega_{2}$. This gives rise to the
\textit{state-independent} inter-layer transitions depicted in Fig.~\ref{Fig1}(c). 
To drive the such transitions, the frequencies
of laser beams are assumed to satisfy the condition of two-photon
interlayer resonance, $\delta\omega_{2}=\Delta_{\mathrm{inter}}$.
The resultant Hamiltonian for the
laser-assisted tunneling takes the form 
\begin{align}
\hat{H}_{{\rm inter}}^{\prime} & =\int d^{2}\mathbf{r}_{_{\bot}}\sum_{\gamma}\left(Je^{i\mathbf{k}_{J}^{\perp}\cdot\mathbf{r}_{_{\bot}}-i\delta\omega_{2}t}+\mathrm{c.c.}\right)\hat{\Phi}_{\gamma2}^{\dag}\hat{\Phi}_{\gamma1}\nonumber \\
 & +{\rm H.c.\,,}\label{eq:H3}
\end{align}
where $\mathbf{k}_{J}=\mathbf{k}_{2}-\mathbf{k}_{0}=k_{J}^{\perp}\mathbf{e}_{+}+k_{J}^{z}\mathbf{e}_{z}$
with $k_{J}^{\perp}=k_{2}$ and $k_{J}^{z}=k_{0}$. Here $J=\Omega_{J}\int dz\phi_{2}^{*}(z)\phi_{1}(z)e^{ik_{J}^{z}z}$
is the inter-layer coupling with $\Omega_{J}$ being the corresponding
Rabi frequency, whereas $\phi_{1,2}(z)$ are the Wannier-like states
localized at the layer $1$ or $2$. Note that the Wannier-like states
$\phi_{1}(z)$ and $\phi_{2}(z)$ are orthogonal. Therefore the non-vanishing
overlap integral determining $J$ comes from the contribution of the
factor $e^{ik_{J}^{z}z}\equiv e^{ik_0z}$ due to the momentum transfer along the tunneling
direction $\mathbf{e}_{z}$ \cite{Miyake2013}. Since the length of
the in-plane wave-vectors $\mathbf{k}_{1}$ and $\mathbf{k}_{2}$
is almost the same, in the following we shall take $k_{J}^{\perp}=k_{\Omega}^{\perp}=\kappa$.

\subsection{Elimination of the spatial and temporal dependence}

To gauge away the spatial and temporal dependence in the atom-light
interaction operators $\hat{H}_{{\rm intra}}^{\prime}$ and $\hat{H}_{{\rm inter}}^{\prime}$,
a fast oscillating (both spatially and temporarily) phase is factored
out from each operator $\hat{\Phi}_{j\gamma}(\mathbf{r}_{_{\bot}})$
by writing 
\begin{equation}
\left(\begin{array}{c}
\hat{\Phi}_{\uparrow1}(\mathbf{r}_{_{\bot}})\\
\hat{\Phi}_{\downarrow1}(\mathbf{r}_{_{\bot}})\\
\hat{\Phi}_{\uparrow2}(\mathbf{r}_{_{\bot}})\\
\hat{\Phi}_{\downarrow2}(\mathbf{r}_{_{\bot}})
\end{array}\right)=\left(\begin{array}{l}
\hat{\psi}_{\uparrow1}(\mathbf{r}_{_{\bot}})e^{-i\kappa y-i\epsilon_{1\uparrow}t}\\
\hat{\psi}_{\downarrow1}(\mathbf{r}_{_{\bot}})e^{-i\kappa x-i\epsilon_{1\downarrow}t}\\
\hat{\psi}_{\uparrow2}(\mathbf{r}_{_{\bot}})e^{i\kappa x-i\epsilon_{2\uparrow}t}\\
\hat{\psi}_{\downarrow2}(\mathbf{r}_{_{\bot}})e^{i\kappa y-i\epsilon_{2\downarrow}t}
\end{array}\right)\,.\label{eq:transformation-1}
\end{equation}
Applying the rotating wave approximation,
the resultant time- and position-independent single-particle Hamiltonian
is given by Eqs.~(\ref{H00})--(\ref{eq:H-SOC})
in the main text. Note that the gauge transformation (\ref{eq:transformation-1})
introduces an additional SOC term $\hat{H}_{\mathrm{SOC}}$ given
by Eq.~(\ref{eq:H-SOC}).

\section{Eigenvalue problem and Hamiltonian in rotated basis}

\subsection{The single-particle Hamiltonian and its eigenstates}

Denoting 
\begin{eqnarray}
|\uparrow,1\rangle & = & \left(\begin{array}{c}
1\\
0\\
0\\
0
\end{array}\right)\,,\quad|\downarrow,1\rangle=\left(\begin{array}{c}
0\\
1\\
0\\
0
\end{array}\right)\,,\label{eq:columns}\\
|\uparrow,2\rangle & = & \left(\begin{array}{c}
0\\
0\\
1\\
0
\end{array}\right)\,,\quad|\downarrow,2\rangle=\left(\begin{array}{c}
0\\
0\\
0\\
1
\end{array}\right)\,,\nonumber 
\end{eqnarray}
the single-particle Hamiltonian, Eqs.~(\ref{H_0})--(\ref{H3-1}),
can be expressed in the momentum space as: 
\begin{equation}
H_{0}=\frac{\hbar^{2}}{2m}(k^{2}+\kappa^{2})+H_{\mathrm{SOC},\mathbf{k}}\,,\label{eq:H-matrix}
\end{equation}
where 
\begin{equation}
H_{\mathrm{SOC},\mathbf{k}}=\frac{\hbar^{2}}{m}\mathbf{k}\cdot\mathbf{q}+H_{\mathrm{inter}}+H_{\mathrm{intra}}\,,\label{eq:H_SOC-matrix}
\end{equation}
and 
\begin{eqnarray}
\mathbf{q} & = & \kappa\mathbf{e}_{x}\left(|\uparrow,2\rangle\langle\uparrow,2|-|\downarrow,1\rangle\langle\downarrow,1|\right)\nonumber \\
 &  & +\kappa\mathbf{e}_{y}\left(|\downarrow,2\rangle\langle\downarrow,2|-|\uparrow,1\rangle\langle\uparrow,1|\right)\,,\\
H_{\mathrm{inter}} & = & J\left(|\uparrow,2\rangle\langle\uparrow,1|+|\downarrow,2\rangle\langle\downarrow,1|\right)+\mathrm{H.c.}\,,\\
H_{\mathrm{intra}} & = & \Omega\left(e^{i\varphi}|\uparrow,1\rangle\langle\downarrow,1|+e^{-i\varphi}|\uparrow,2\rangle\langle\downarrow,2|\right)\nonumber \\
 &  & +\mathrm{H.c.}\,.
\end{eqnarray}
where the momentum $\mathbf{k}\equiv \mathbf{k}_\bot = (k_x, k_y)$ is in the $xy$ plane.

The Hamiltonian $H_{\mathrm{SOC},\mathbf{k}}$ can be represented
in a block diagonal form: 
\begin{equation}
H_{\mathrm{SOC},\mathbf{k}}=\left(\begin{array}{cc}
h_{1,\mathbf{k}} & J\\
J & h_{2,\mathbf{k}}
\end{array}\right)\label{eq:H_SOC-matrix-explicit}
\end{equation}
with 
\begin{equation}
h_{1,\mathbf{k}}=\left(\begin{array}{cc}
-2k_{y} & \Omega e^{i\varphi}\\
\Omega e^{-i\varphi} & -2k_{x}
\end{array}\right)\,,\quad h_{2,\mathbf{k}}=\left(\begin{array}{cc}
2k_{x} & \Omega e^{-i\varphi}\\
\Omega e^{i\varphi} & 2k_{y}
\end{array}\right)\,,\label{eq:h1-h2}
\end{equation}
and the off-diagonal $2\times2$ blocks $J\equiv JI$ being proportional
to the $2\times2$ unit matrix $I$. The block diagonal form
of the Hamiltonian $H_{0}$ given by Eqs. (\ref{eq:H-matrix}),
(\ref{eq:H_SOC-matrix-explicit}) and (\ref{eq:h1-h2}), allows
to find its eigenstates in a straightforward way: 
\begin{equation}
E_{\alpha,\eta}=1+k^{2}+\alpha\sqrt{\Omega^{2}+J^{2}+2k^{2}+2\eta a_{\mathbf{k}}}\,,
\end{equation}
with $\alpha=\pm1$, $\eta=\pm1$, where the energy is measured
in the units of the of the recoil energy $E_{\mathrm{rec}}=\hbar^{2}\kappa^{2}/{2m}$,
whereas the momentum is measured in terms of the recoil
momentum $\kappa$. Here 
\begin{equation}
a_{\mathbf{k}}=\sqrt{\Omega^{2}(k_{x}+k_{y})^{2}+J^{2}(k_{x}-k_{y})^{2}+(k_{x}^{2}-k_{y}^{2})^{2}}\,,
\end{equation}
with $\mathbf{k}= k(\cos\phi\,\mathbf{e}_{x}+\sin\phi\,\mathbf{e}_{y})/2$,
and $\phi$ being an azimuthal angle in the momentum space. 
The lowest dispersion branch 
\begin{equation}
E_{g}=E_{-1,1}=1+k^{2}-\sqrt{\Omega^{2}+J^{2}+2k^{2}+2a_{\mathbf{k}}}\,,\label{eq:ground-state-dis-1}
\end{equation}
is obtained by taking $\alpha=-1$, $\eta=+1$. For $\Omega^2=J^2\gg E_{\mathrm{rec}}^2$
the eigenvector corresponding to the lowest dispersion branch is given
by Eq.~(\ref{eq:eigenvector-ring}) of the main text: 
\begin{equation}
\chi=\left(\begin{array}{c}
\sqrt{2}\cos\phi\\
i(1-\sin\phi+\cos\phi)\\
1-\sin\phi-\cos\phi\\
-\sqrt{2}i(1-\sin\phi)
\end{array}\right)\frac{e^{i\mathbf{k}\cdot\mathbf{r}_{_{\bot}}}}{\sqrt{8-8\sin\phi}}\,.\label{eq:eigenvector-ring-1}
\end{equation}

To gain more insight into the interaction-induced
symmetry breaking, we will present the Hamiltonian in terms of the basis vectors for
which the atoms possess the minimum interaction energy on the Rashba
ring. The projection of such a Hamiltonian to lower energy states
gives rise to the appearance of Rashba Hamiltonian. These issues will be addressed next.

\subsection{Hamiltonian in rotated basis and reduction to the Rashba Hamiltonian}

In Sec.~III~C of the main text, the variational approach shows
that the minimization of interaction energy breaks the rotational
symmetry of the Rashba ring. 
In the case of single momentum states (PW-II phase), the interaction
energy between the atoms acquires a minimum value for $\phi=3\pi/4$
or $\phi=-\pi/4$, i.e. for $\mathbf{k}$ and $\mathbf{-k}$ along a diagonal $\mathbf{e}_{x} - \mathbf{e}_{y}$. For these
azimuthal angles the spinor part of eigenvectors $\chi^{(1)}=\chi(3\pi/4)$
and $\chi^{(2)}=\chi(-\pi/4)$ read using Eq.(\ref{eq:eigenvector-ring-1})
or Eq.~(\ref{eq:eigenvector-ring}) in the main text:

\begin{equation}
\chi^{(1)}=\frac{1}{b_{-}}\left(\begin{array}{c}
-1\\
ia_{-}\\
1\\
ia_{-}
\end{array}\right)\,,\qquad\chi^{(2)}=\frac{1}{b_{+}}\left(\begin{array}{c}
1\\
ia_{+}\\
1\\
-ia_{+}
\end{array}\right)\,,\label{eq:rotated-basis-low-energy}
\end{equation}
where 
\[
b_{\pm}=2\sqrt{2\pm\sqrt{2}}\,,\qquad a_{\pm}=1\pm\sqrt{2}\,.
\]
The vectors $\chi^{(1)}$ and $\chi^{(2)}$ can serve as a basis for
the lowest dispersion branch. To have a complete rotated bases, we
choose the remaining two orthogonal vectors to be 
\begin{equation}
\chi^{(3)}=\frac{1}{b_{-}}\left(\begin{array}{c}
-1\\
i\\
a_{-}\\
-ia_{-}
\end{array}\right)\,,\qquad\chi^{(4)}=\frac{1}{b_{+}}\left(\begin{array}{c}
1\\
i\\
a_{+}\\
ia_{+}
\end{array}\right)\,.\label{eq:rotated-basis-high-energy}
\end{equation}
In the rotated basis, the Hamiltonian $H_{\mathrm{SOC},\mathbf{k}}$,
Eq.~(\ref{eq:H_SOC-matrix-explicit}), reads for $\Omega = J$ 
\begin{align}
H_{\mathrm{SOC},\mathbf{k}} & =\left(\begin{array}{cc}
-\sqrt{2}\Omega & 0\\
0 & \sqrt{2}\Omega
\end{array}\right)\nonumber \\
 & +\left(\begin{array}{cc}
\frac{1}{\sqrt{2}}(k_{-}\sigma_{z}+k_{+}\sigma_{x}) & -k_{y}I+ik_{x}\sigma_{y}\\
-k_{y}I-ik_{x}\sigma_{y} & -\frac{1}{\sqrt{2}}(k_{-}\sigma_{x}+k_{+}\sigma_{z})
\end{array}\right)\,,\label{eq:H_SOC-matrix-rotated-basis}
\end{align}
with $k_{\pm}=k_{x}\pm k_{y}$.

For $\Omega^2=J^2\gg E_{\mathrm{rec}}^2$ , the upper and lower pairs states
are separated by the energy $\approx2\sqrt{2}\Omega$. In that case
one can neglect the coupling between the lower and upper two pairs
of states. The Hamiltonian projected onto the manifold of low-energy
states $\chi^{(1)}$ and $\chi^{(2)}$ reduces to the usual Rashba-type
Hamiltonian 
\begin{equation}
H_{\mathrm{SOC},\mathbf{k}}\rightarrow\frac{1}{\sqrt{2}}(k_{-}\sigma_{z}+k_{+}\sigma_{x})\,.\label{eq:H-Rashba}
\end{equation}
subject to the rotation of the spin by $\pi/2$ along the $x$ axis
transforming $\sigma_{z}$ to $\sigma_{y}$.

\subsection{Interaction energy}

According to the interaction Hamiltonian (\ref{intH}), the contact
interaction between atoms is described by the functional 
\begin{equation}
E_{\mathrm{int}}=\sum_{j=1}^{2}\int d^{2}\mathbf{r}\,\left(\frac{g_{\uparrow}}{2}|\psi_{\uparrow,j}|^{4}+\frac{g_{\downarrow}}{2}|\psi_{\downarrow,j}|^{4}+g_{\uparrow\downarrow}|\psi_{\uparrow,j}|^{2}|\psi_{\downarrow,j}|^{2}\right)\,.\label{eq:inter-energy-app-b}
\end{equation}
Let us assume that the state of the atomic cloud is a superposition
of lowest states $\chi^{(1)}$ and $\chi^{(2)}$ with the coefficients
$\tilde{\psi}_{1}$ and $\tilde{\psi}_{2}$: 
\begin{equation}
\chi=\chi^{(1)}\tilde{\psi}_{1}+\chi^{(2)}\tilde{\psi}_{2}\label{eq:chi-superposition}
\end{equation}
For $g_{\uparrow}=g_{\downarrow}$, the interaction energy (\ref{eq:inter-energy-app-b})
becomes 
\begin{align}
E_{\mathrm{int}} & =\frac{1}{16}\left[(3g_{\uparrow}+g_{\uparrow\downarrow})(|\tilde{\psi}_{1}|^{4}+|\tilde{\psi}_{2}|^{4})+4(g_{\uparrow}+2g_{\uparrow\downarrow})|\tilde{\psi}_{1}|^{2}|\tilde{\psi}_{2}|^{2}\right.\nonumber \\
 & +\left.(g_{\uparrow}+g_{\uparrow\downarrow})\left(\tilde{\psi}_{1}^{*2}\tilde{\psi}_{2}^{2}+\tilde{\psi}_{2}^{*2}\tilde{\psi}_{1}^{2}\right)\right].\label{eq:e-2}
\end{align}
Taking $\tilde{\psi}_{1}=-\psi\sin(\frac{1}{2}(\phi+\frac{\pi}{4}))$,
$\tilde{\psi}_{2}=\psi\cos(\frac{1}{2}(\phi+\frac{\pi}{4}))$ the
superposition vector (\ref{eq:chi-superposition}) reduces to Eq.(\ref{eq:eigenvector-ring-1})
or Eq.~(\ref{eq:eigenvector-ring}) in the main text. For $\phi=-\pi/4$, one has $\tilde{\psi}_{1}=0$,
so there is only one column $\chi^{(2)}$. Thus $\phi$ indeed represents
the azimuthal angle in the momentum space. Inserting the expressions
for $\tilde{\psi}_{1}$ and $\tilde{\psi}_{2}$ into Eq.~(\ref{eq:e-2})
we get 
\begin{equation}
E_{\mathrm{int}}=\frac{c_{0}}{8}+\frac{c_{0}}{32}\left(1+\sin2\phi\right)+\frac{c_{2}}{32}\left(1-\sin2\phi\right)\,\label{eq:E_int-rotated-basis}
\end{equation}
where we assume $g_{\uparrow}=g_{\downarrow}=g$ and introduce $c_{0}=g+g_{\uparrow\downarrow}$
and $c_{2}=g-g_{\uparrow\downarrow}$, with $c_{0}\gg c_{2}$. This
is equivalent to the expression (\ref{eq:E_int_PW_II}) of the main
text for the PW-II phase. Using the projected basis $\chi_{1}$ and
$\chi_{2}$, the interaction energy $E_{\mathrm{int}}$ given by Eq.~(\ref{eq:E_int-rotated-basis})
acquires a minimum value for $\phi=3\pi/4$ or $\phi=-\pi/4$,
as required. Thus the interaction appears to be highly anisotropic
along the Rashba ring.

%\bibliography{SOC_bilayer_c}

\begin{thebibliography}{70}%
\makeatletter
\providecommand \@ifxundefined [1]{%
 \@ifx{#1\undefined}
}%
\providecommand \@ifnum [1]{%
 \ifnum #1\expandafter \@firstoftwo
 \else \expandafter \@secondoftwo
 \fi
}%
\providecommand \@ifx [1]{%
 \ifx #1\expandafter \@firstoftwo
 \else \expandafter \@secondoftwo
 \fi
}%
\providecommand \natexlab [1]{#1}%
\providecommand \enquote  [1]{``#1''}%
\providecommand \bibnamefont  [1]{#1}%
\providecommand \bibfnamefont [1]{#1}%
\providecommand \citenamefont [1]{#1}%
\providecommand \href@noop [0]{\@secondoftwo}%
\providecommand \href [0]{\begingroup \@sanitize@url \@href}%
\providecommand \@href[1]{\@@startlink{#1}\@@href}%
\providecommand \@@href[1]{\endgroup#1\@@endlink}%
\providecommand \@sanitize@url [0]{\catcode `\\12\catcode `\$12\catcode
  `\&12\catcode `\#12\catcode `\^12\catcode `\_12\catcode `\%12\relax}%
\providecommand \@@startlink[1]{}%
\providecommand \@@endlink[0]{}%
\providecommand \url  [0]{\begingroup\@sanitize@url \@url }%
\providecommand \@url [1]{\endgroup\@href {#1}{\urlprefix }}%
\providecommand \urlprefix  [0]{URL }%
\providecommand \Eprint [0]{\href }%
\providecommand \doibase [0]{http://dx.doi.org/}%
\providecommand \selectlanguage [0]{\@gobble}%
\providecommand \bibinfo  [0]{\@secondoftwo}%
\providecommand \bibfield  [0]{\@secondoftwo}%
\providecommand \translation [1]{[#1]}%
\providecommand \BibitemOpen [0]{}%
\providecommand \bibitemStop [0]{}%
\providecommand \bibitemNoStop [0]{.\EOS\space}%
\providecommand \EOS [0]{\spacefactor3000\relax}%
\providecommand \BibitemShut  [1]{\csname bibitem#1\endcsname}%
\let\auto@bib@innerbib\@empty
%</preamble>
\bibitem [{\citenamefont {Lin}\ \emph {et~al.}(2009)\citenamefont {Lin},
  \citenamefont {Compton}, \citenamefont {Jim\'{e}nez-Garc\'{i}a},
  \citenamefont {Porto},\ and\ \citenamefont {Spielman}}]{Lin2009}%
  \BibitemOpen
  \bibfield  {author} {\bibinfo {author} {\bibfnamefont {Y.-J.}\ \bibnamefont
  {Lin}}, \bibinfo {author} {\bibfnamefont {R.~L.}\ \bibnamefont {Compton}},
  \bibinfo {author} {\bibfnamefont {K.}~\bibnamefont {Jim\'{e}nez-Garc\'{i}a}},
  \bibinfo {author} {\bibfnamefont {J.~V.}\ \bibnamefont {Porto}}, \ and\
  \bibinfo {author} {\bibfnamefont {I.~B.}\ \bibnamefont {Spielman}},\
  }\href@noop {} {\bibfield  {journal} {\bibinfo  {journal} {Nature (London)}\
  }\textbf {\bibinfo {volume} {462}},\ \bibinfo {pages} {628} (\bibinfo {year}
  {2009})}\BibitemShut {NoStop}%
\bibitem [{\citenamefont {Struck}\ \emph {et~al.}(2012)\citenamefont {Struck},
  \citenamefont {{\"O}lschl{\"a}ger}, \citenamefont {Weinberg}, \citenamefont
  {Hauke}, \citenamefont {Simonet}, \citenamefont {Eckardt}, \citenamefont
  {Lewenstein}, \citenamefont {Sengstock},\ and\ \citenamefont
  {Windpassinger}}]{Struck2012}%
  \BibitemOpen
  \bibfield  {author} {\bibinfo {author} {\bibfnamefont {J.}~\bibnamefont
  {Struck}}, \bibinfo {author} {\bibfnamefont {C.}~\bibnamefont
  {{\"O}lschl{\"a}ger}}, \bibinfo {author} {\bibfnamefont {M.}~\bibnamefont
  {Weinberg}}, \bibinfo {author} {\bibfnamefont {P.}~\bibnamefont {Hauke}},
  \bibinfo {author} {\bibfnamefont {J.}~\bibnamefont {Simonet}}, \bibinfo
  {author} {\bibfnamefont {A.}~\bibnamefont {Eckardt}}, \bibinfo {author}
  {\bibfnamefont {M.}~\bibnamefont {Lewenstein}}, \bibinfo {author}
  {\bibfnamefont {K.}~\bibnamefont {Sengstock}}, \ and\ \bibinfo {author}
  {\bibfnamefont {P.}~\bibnamefont {Windpassinger}},\ }\href@noop {} {\bibfield
   {journal} {\bibinfo  {journal} {Phys. Rev. Lett.}\ }\textbf {\bibinfo
  {volume} {108}},\ \bibinfo {pages} {225304} (\bibinfo {year}
  {2012})}\BibitemShut {NoStop}%
\bibitem [{\citenamefont {Aidelsburger}\ \emph {et~al.}(2013)\citenamefont
  {Aidelsburger}, \citenamefont {Atala}, \citenamefont {Lohse}, \citenamefont
  {Barreiro}, \citenamefont {Paredes},\ and\ \citenamefont
  {Bloch}}]{Aidelsburger2013}%
  \BibitemOpen
  \bibfield  {author} {\bibinfo {author} {\bibfnamefont {M.}~\bibnamefont
  {Aidelsburger}}, \bibinfo {author} {\bibfnamefont {M.}~\bibnamefont {Atala}},
  \bibinfo {author} {\bibfnamefont {M.}~\bibnamefont {Lohse}}, \bibinfo
  {author} {\bibfnamefont {J.~T.}\ \bibnamefont {Barreiro}}, \bibinfo {author}
  {\bibfnamefont {B.}~\bibnamefont {Paredes}}, \ and\ \bibinfo {author}
  {\bibfnamefont {I.}~\bibnamefont {Bloch}},\ }\href {\doibase
  10.1103/PhysRevLett.111.185301} {\bibfield  {journal} {\bibinfo  {journal}
  {Physical Review Letters}\ }\textbf {\bibinfo {volume} {111}},\ \bibinfo
  {pages} {185301} (\bibinfo {year} {2013})}\BibitemShut {NoStop}%
\bibitem [{\citenamefont {Miyake}\ \emph {et~al.}(2013)\citenamefont {Miyake},
  \citenamefont {Siviloglou}, \citenamefont {Kennedy}, \citenamefont {Burton},\
  and\ \citenamefont {Ketterle}}]{Miyake2013}%
  \BibitemOpen
  \bibfield  {author} {\bibinfo {author} {\bibfnamefont {H.}~\bibnamefont
  {Miyake}}, \bibinfo {author} {\bibfnamefont {G.~A.}\ \bibnamefont
  {Siviloglou}}, \bibinfo {author} {\bibfnamefont {C.~J.}\ \bibnamefont
  {Kennedy}}, \bibinfo {author} {\bibfnamefont {W.~C.}\ \bibnamefont {Burton}},
  \ and\ \bibinfo {author} {\bibfnamefont {W.}~\bibnamefont {Ketterle}},\
  }\href {\doibase 10.1103/PhysRevLett.111.185302} {\bibfield  {journal}
  {\bibinfo  {journal} {Phys. Rev. Lett.}\ }\textbf {\bibinfo {volume} {111}},\
  \bibinfo {pages} {185302} (\bibinfo {year} {2013})}\BibitemShut {NoStop}%
\bibitem [{\citenamefont {Jotzu}\ \emph {et~al.}(2014)\citenamefont {Jotzu},
  \citenamefont {Messer}, \citenamefont {Desbuquois}, \citenamefont {Lebrat},
  \citenamefont {Uehlinger}, \citenamefont {Greif},\ and\ \citenamefont
  {Esslinger}}]{Esslinger2014}%
  \BibitemOpen
  \bibfield  {author} {\bibinfo {author} {\bibfnamefont {G.}~\bibnamefont
  {Jotzu}}, \bibinfo {author} {\bibfnamefont {M.}~\bibnamefont {Messer}},
  \bibinfo {author} {\bibfnamefont {R.}~\bibnamefont {Desbuquois}}, \bibinfo
  {author} {\bibfnamefont {M.}~\bibnamefont {Lebrat}}, \bibinfo {author}
  {\bibfnamefont {T.}~\bibnamefont {Uehlinger}}, \bibinfo {author}
  {\bibfnamefont {D.}~\bibnamefont {Greif}}, \ and\ \bibinfo {author}
  {\bibfnamefont {T.}~\bibnamefont {Esslinger}},\ }\href@noop {} {\bibfield
  {journal} {\bibinfo  {journal} {Nature}\ }\textbf {\bibinfo {volume} {515}},\
  \bibinfo {pages} {237} (\bibinfo {year} {2014})}\BibitemShut {NoStop}%
\bibitem [{\citenamefont {Dum}\ and\ \citenamefont {Olshanii}(1996)}]{Dum1996}%
  \BibitemOpen
  \bibfield  {author} {\bibinfo {author} {\bibfnamefont {R.}~\bibnamefont
  {Dum}}\ and\ \bibinfo {author} {\bibfnamefont {M.}~\bibnamefont {Olshanii}},\
  }\href@noop {} {\bibfield  {journal} {\bibinfo  {journal} {Phys. Rev. Lett.}\
  }\textbf {\bibinfo {volume} {76}},\ \bibinfo {pages} {1788} (\bibinfo {year}
  {1996})}\BibitemShut {NoStop}%
\bibitem [{\citenamefont {Visser}\ and\ \citenamefont
  {Nienhuis}(1998)}]{Visser1998}%
  \BibitemOpen
  \bibfield  {author} {\bibinfo {author} {\bibfnamefont {P.~M.}\ \bibnamefont
  {Visser}}\ and\ \bibinfo {author} {\bibfnamefont {G.}~\bibnamefont
  {Nienhuis}},\ }\href {\doibase 10.1103/PhysRevA.57.4581} {\bibfield
  {journal} {\bibinfo  {journal} {Phys. Rev. A}\ }\textbf {\bibinfo {volume}
  {57}},\ \bibinfo {pages} {4581} (\bibinfo {year} {1998})}\BibitemShut
  {NoStop}%
\bibitem [{\citenamefont {Juzeli\=unas}\ and\ \citenamefont
  {\"Ohberg}(2004)}]{Juzeliunas2004}%
  \BibitemOpen
  \bibfield  {author} {\bibinfo {author} {\bibfnamefont {G.}~\bibnamefont
  {Juzeli\=unas}}\ and\ \bibinfo {author} {\bibfnamefont {P.}~\bibnamefont
  {\"Ohberg}},\ }\href@noop {} {\bibfield  {journal} {\bibinfo  {journal}
  {Phys. Rev. Lett.}\ }\textbf {\bibinfo {volume} {93}},\ \bibinfo {pages}
  {033602} (\bibinfo {year} {2004})}\BibitemShut {NoStop}%
\bibitem [{\citenamefont {Juzeli{\=u}nas}\ \emph {et~al.}(2006)\citenamefont
  {Juzeli{\=u}nas}, \citenamefont {Ruseckas}, \citenamefont {{\"O}hberg},\ and\
  \citenamefont {Fleischhauer}}]{Juzeliunas2006}%
  \BibitemOpen
  \bibfield  {author} {\bibinfo {author} {\bibfnamefont {G.}~\bibnamefont
  {Juzeli{\=u}nas}}, \bibinfo {author} {\bibfnamefont {J.}~\bibnamefont
  {Ruseckas}}, \bibinfo {author} {\bibfnamefont {P.}~\bibnamefont
  {{\"O}hberg}}, \ and\ \bibinfo {author} {\bibfnamefont {M.}~\bibnamefont
  {Fleischhauer}},\ }\href@noop {} {\bibfield  {journal} {\bibinfo  {journal}
  {Phys. Rev. A}\ }\textbf {\bibinfo {volume} {73}},\ \bibinfo {pages} {025602}
  (\bibinfo {year} {2006})}\BibitemShut {NoStop}%
\bibitem [{\citenamefont {Dalibard}\ \emph {et~al.}(2011)\citenamefont
  {Dalibard}, \citenamefont {Gerbier}, \citenamefont {Juzeli{\=u}nas},\ and\
  \citenamefont {{\"O}hberg}}]{Dalibard2011}%
  \BibitemOpen
  \bibfield  {author} {\bibinfo {author} {\bibfnamefont {J.}~\bibnamefont
  {Dalibard}}, \bibinfo {author} {\bibfnamefont {F.}~\bibnamefont {Gerbier}},
  \bibinfo {author} {\bibfnamefont {G.}~\bibnamefont {Juzeli{\=u}nas}}, \ and\
  \bibinfo {author} {\bibfnamefont {P.}~\bibnamefont {{\"O}hberg}},\
  }\href@noop {} {\bibfield  {journal} {\bibinfo  {journal} {Rev. Mod. Phys.}\
  }\textbf {\bibinfo {volume} {83}},\ \bibinfo {pages} {1523} (\bibinfo {year}
  {2011})}\BibitemShut {NoStop}%
\bibitem [{\citenamefont {Goldman}\ \emph {et~al.}(2014)\citenamefont
  {Goldman}, \citenamefont {Juzeli{\=u}nas}, \citenamefont {{\"O}hberg},\ and\
  \citenamefont {Spielman}}]{Goldman2014}%
  \BibitemOpen
  \bibfield  {author} {\bibinfo {author} {\bibfnamefont {N.}~\bibnamefont
  {Goldman}}, \bibinfo {author} {\bibfnamefont {G.}~\bibnamefont
  {Juzeli{\=u}nas}}, \bibinfo {author} {\bibfnamefont {P.}~\bibnamefont
  {{\"O}hberg}}, \ and\ \bibinfo {author} {\bibfnamefont {I.~B.}\ \bibnamefont
  {Spielman}},\ }\href@noop {} {\bibfield  {journal} {\bibinfo  {journal} {Rep.
  Progr. Phys.}\ }\textbf {\bibinfo {volume} {77}},\ \bibinfo {pages} {126401}
  (\bibinfo {year} {2014})}\BibitemShut {NoStop}%
\bibitem [{\citenamefont {Lewenstein}\ \emph {et~al.}(2012)\citenamefont
  {Lewenstein}, \citenamefont {Anna},\ and\ \citenamefont
  {Ver{\`o}nica}}]{Lewenstein2012}%
  \BibitemOpen
  \bibfield  {author} {\bibinfo {author} {\bibfnamefont {M.}~\bibnamefont
  {Lewenstein}}, \bibinfo {author} {\bibfnamefont {S.}~\bibnamefont {Anna}}, \
  and\ \bibinfo {author} {\bibfnamefont {A.}~\bibnamefont {Ver{\`o}nica}},\
  }\href@noop {} {\emph {\bibinfo {title} {Ultracold Atoms in Optical Lattices:
  Simulating quantum many-body systems}}}\ (\bibinfo  {publisher} {Oxford
  University Press},\ \bibinfo {year} {2012})\BibitemShut {NoStop}%
\bibitem [{\citenamefont {Ruseckas}\ \emph {et~al.}(2005)\citenamefont
  {Ruseckas}, \citenamefont {Juzeli{\=u}nas}, \citenamefont {{\"O}hberg},\ and\
  \citenamefont {Fleischhauer}}]{Ruseckas2005}%
  \BibitemOpen
  \bibfield  {author} {\bibinfo {author} {\bibfnamefont {J.}~\bibnamefont
  {Ruseckas}}, \bibinfo {author} {\bibfnamefont {G.}~\bibnamefont
  {Juzeli{\=u}nas}}, \bibinfo {author} {\bibfnamefont {P.}~\bibnamefont
  {{\"O}hberg}}, \ and\ \bibinfo {author} {\bibfnamefont {M.}~\bibnamefont
  {Fleischhauer}},\ }\href@noop {} {\bibfield  {journal} {\bibinfo  {journal}
  {Phys. Rev. Lett.}\ }\textbf {\bibinfo {volume} {95}},\ \bibinfo {pages}
  {010404} (\bibinfo {year} {2005})}\BibitemShut {NoStop}%
\bibitem [{\citenamefont {Zhai}(2015)}]{Zhai2014-review}%
  \BibitemOpen
  \bibfield  {author} {\bibinfo {author} {\bibfnamefont {H.}~\bibnamefont
  {Zhai}},\ }\href@noop {} {\bibfield  {journal} {\bibinfo  {journal} {Rep.
  Prog. Phys.}\ }\textbf {\bibinfo {volume} {78}},\ \bibinfo {pages} {026001}
  (\bibinfo {year} {2015})}\BibitemShut {NoStop}%
\bibitem [{\citenamefont {Wang}\ \emph {et~al.}(2010)\citenamefont {Wang},
  \citenamefont {Gao}, \citenamefont {Jian},\ and\ \citenamefont
  {Zhai}}]{Wang2010}%
  \BibitemOpen
  \bibfield  {author} {\bibinfo {author} {\bibfnamefont {C.}~\bibnamefont
  {Wang}}, \bibinfo {author} {\bibfnamefont {C.}~\bibnamefont {Gao}}, \bibinfo
  {author} {\bibfnamefont {C.-M.}\ \bibnamefont {Jian}}, \ and\ \bibinfo
  {author} {\bibfnamefont {H.}~\bibnamefont {Zhai}},\ }\href@noop {} {\bibfield
   {journal} {\bibinfo  {journal} {Phys. Rev. Lett.}\ }\textbf {\bibinfo
  {volume} {105}},\ \bibinfo {pages} {160403} (\bibinfo {year}
  {2010})}\BibitemShut {NoStop}%
\bibitem [{\citenamefont {Xu}\ \emph {et~al.}(2011)\citenamefont {Xu},
  \citenamefont {L\"{u}},\ and\ \citenamefont {You}}]{Xu2011}%
  \BibitemOpen
  \bibfield  {author} {\bibinfo {author} {\bibfnamefont {Z.~F.}\ \bibnamefont
  {Xu}}, \bibinfo {author} {\bibfnamefont {R.}~\bibnamefont {L\"{u}}}, \ and\
  \bibinfo {author} {\bibfnamefont {L.}~\bibnamefont {You}},\ }\href@noop {}
  {\bibfield  {journal} {\bibinfo  {journal} {Phys. Rev. A}\ }\textbf {\bibinfo
  {volume} {83}},\ \bibinfo {pages} {053602} (\bibinfo {year}
  {2011})}\BibitemShut {NoStop}%
\bibitem [{\citenamefont {Liu}\ and\ \citenamefont {Liu}(2012)}]{Liu2012PRA}%
  \BibitemOpen
  \bibfield  {author} {\bibinfo {author} {\bibfnamefont {C.-F.}\ \bibnamefont
  {Liu}}\ and\ \bibinfo {author} {\bibfnamefont {W.~M.}\ \bibnamefont {Liu}},\
  }\href@noop {} {\bibfield  {journal} {\bibinfo  {journal} {Phys. Rev. A}\
  }\textbf {\bibinfo {volume} {86}},\ \bibinfo {pages} {033602} (\bibinfo
  {year} {2012})}\BibitemShut {NoStop}%
\bibitem [{\citenamefont {Radi{\'c}}\ \emph {et~al.}(2011)\citenamefont
  {Radi{\'c}}, \citenamefont {Sedrakyan}, \citenamefont {Spielman},\ and\
  \citenamefont {Galitski}}]{Radic2011}%
  \BibitemOpen
  \bibfield  {author} {\bibinfo {author} {\bibfnamefont {J.}~\bibnamefont
  {Radi{\'c}}}, \bibinfo {author} {\bibfnamefont {T.~A.}\ \bibnamefont
  {Sedrakyan}}, \bibinfo {author} {\bibfnamefont {I.~B.}\ \bibnamefont
  {Spielman}}, \ and\ \bibinfo {author} {\bibfnamefont {V.}~\bibnamefont
  {Galitski}},\ }\href@noop {} {\bibfield  {journal} {\bibinfo  {journal}
  {Phys. Rev. A}\ }\textbf {\bibinfo {volume} {84}},\ \bibinfo {pages} {063604}
  (\bibinfo {year} {2011})}\BibitemShut {NoStop}%
\bibitem [{\citenamefont {Zhou}\ \emph
  {et~al.}(2011{\natexlab{a}})\citenamefont {Zhou}, \citenamefont {Zhou},\ and\
  \citenamefont {Wu}}]{Zhou2011}%
  \BibitemOpen
  \bibfield  {author} {\bibinfo {author} {\bibfnamefont {X.-F.}\ \bibnamefont
  {Zhou}}, \bibinfo {author} {\bibfnamefont {J.}~\bibnamefont {Zhou}}, \ and\
  \bibinfo {author} {\bibfnamefont {C.}~\bibnamefont {Wu}},\ }\href@noop {}
  {\bibfield  {journal} {\bibinfo  {journal} {Phys. Rev. A}\ }\textbf {\bibinfo
  {volume} {84}},\ \bibinfo {pages} {063624} (\bibinfo {year}
  {2011}{\natexlab{a}})}\BibitemShut {NoStop}%
\bibitem [{\citenamefont {Sinha}\ \emph {et~al.}(2011)\citenamefont {Sinha},
  \citenamefont {Nath},\ and\ \citenamefont {Santos}}]{Sinha2011}%
  \BibitemOpen
  \bibfield  {author} {\bibinfo {author} {\bibfnamefont {S.}~\bibnamefont
  {Sinha}}, \bibinfo {author} {\bibfnamefont {R.}~\bibnamefont {Nath}}, \ and\
  \bibinfo {author} {\bibfnamefont {L.}~\bibnamefont {Santos}},\ }\href@noop {}
  {\bibfield  {journal} {\bibinfo  {journal} {Phys. Rev. Lett.}\ }\textbf
  {\bibinfo {volume} {107}},\ \bibinfo {pages} {270401} (\bibinfo {year}
  {2011})}\BibitemShut {NoStop}%
\bibitem [{\citenamefont {Ozawa}\ and\ \citenamefont
  {Baym}(2012{\natexlab{a}})}]{Ozawa2012a}%
  \BibitemOpen
  \bibfield  {author} {\bibinfo {author} {\bibfnamefont {T.}~\bibnamefont
  {Ozawa}}\ and\ \bibinfo {author} {\bibfnamefont {G.}~\bibnamefont {Baym}},\
  }\href {\doibase 10.1103/PhysRevA.85.013612} {\bibfield  {journal} {\bibinfo
  {journal} {Phys. Rev. A}\ }\textbf {\bibinfo {volume} {85}},\ \bibinfo
  {pages} {013612} (\bibinfo {year} {2012}{\natexlab{a}})}\BibitemShut
  {NoStop}%
\bibitem [{\citenamefont {Ozawa}\ and\ \citenamefont
  {Baym}(2012{\natexlab{b}})}]{Ozawa2012b}%
  \BibitemOpen
  \bibfield  {author} {\bibinfo {author} {\bibfnamefont {T.}~\bibnamefont
  {Ozawa}}\ and\ \bibinfo {author} {\bibfnamefont {G.}~\bibnamefont {Baym}},\
  }\href {\doibase 10.1103/PhysRevA.85.063623} {\bibfield  {journal} {\bibinfo
  {journal} {Phys. Rev. A}\ }\textbf {\bibinfo {volume} {85}},\ \bibinfo
  {pages} {063623} (\bibinfo {year} {2012}{\natexlab{b}})}\BibitemShut
  {NoStop}%
\bibitem [{\citenamefont {Liu}\ \emph {et~al.}(2013)\citenamefont {Liu},
  \citenamefont {Yu}, \citenamefont {Gou},\ and\ \citenamefont
  {Liu}}]{Liu2013PRA}%
  \BibitemOpen
  \bibfield  {author} {\bibinfo {author} {\bibfnamefont {C.-F.}\ \bibnamefont
  {Liu}}, \bibinfo {author} {\bibfnamefont {Y.-M.}\ \bibnamefont {Yu}},
  \bibinfo {author} {\bibfnamefont {S.-C.}\ \bibnamefont {Gou}}, \ and\
  \bibinfo {author} {\bibfnamefont {W.-M.}\ \bibnamefont {Liu}},\ }\href@noop
  {} {\bibfield  {journal} {\bibinfo  {journal} {Phys. Rev. A}\ }\textbf
  {\bibinfo {volume} {87}},\ \bibinfo {pages} {063630} (\bibinfo {year}
  {2013})}\BibitemShut {NoStop}%
\bibitem [{\citenamefont {Chen}\ \emph {et~al.}(2014)\citenamefont {Chen},
  \citenamefont {Rabinovic}, \citenamefont {Anderson},\ and\ \citenamefont
  {Santos}}]{Chen2014PRA}%
  \BibitemOpen
  \bibfield  {author} {\bibinfo {author} {\bibfnamefont {X.}~\bibnamefont
  {Chen}}, \bibinfo {author} {\bibfnamefont {M.}~\bibnamefont {Rabinovic}},
  \bibinfo {author} {\bibfnamefont {B.~M.}\ \bibnamefont {Anderson}}, \ and\
  \bibinfo {author} {\bibfnamefont {L.}~\bibnamefont {Santos}},\ }\href@noop {}
  {\bibfield  {journal} {\bibinfo  {journal} {Phys. Rev. A}\ }\textbf {\bibinfo
  {volume} {90}},\ \bibinfo {pages} {043632} (\bibinfo {year}
  {2014})}\BibitemShut {NoStop}%
\bibitem [{\citenamefont {Han}\ \emph {et~al.}(2015)\citenamefont {Han},
  \citenamefont {Juzeli{\=u}nas}, \citenamefont {Zhang},\ and\ \citenamefont
  {Liu}}]{Han2015}%
  \BibitemOpen
  \bibfield  {author} {\bibinfo {author} {\bibfnamefont {W.}~\bibnamefont
  {Han}}, \bibinfo {author} {\bibfnamefont {G.}~\bibnamefont {Juzeli{\=u}nas}},
  \bibinfo {author} {\bibfnamefont {W.}~\bibnamefont {Zhang}}, \ and\ \bibinfo
  {author} {\bibfnamefont {W.-M.}\ \bibnamefont {Liu}},\ }\href@noop {}
  {\bibfield  {journal} {\bibinfo  {journal} {Phys. Rev. A}\ }\textbf {\bibinfo
  {volume} {91}},\ \bibinfo {pages} {013607} (\bibinfo {year}
  {2015})}\BibitemShut {NoStop}%
\bibitem [{\citenamefont {Su}\ \emph {et~al.}(2015)\citenamefont {Su},
  \citenamefont {Gou}, \citenamefont {Liu}, \citenamefont {Spielman},
  \citenamefont {Santos}, \citenamefont {Acus}, \citenamefont {Mekys},
  \citenamefont {Ruseckas},\ and\ \citenamefont {Juzeli{\=u}nas}}]{Su2015}%
  \BibitemOpen
  \bibfield  {author} {\bibinfo {author} {\bibfnamefont {S.-W.}\ \bibnamefont
  {Su}}, \bibinfo {author} {\bibfnamefont {S.-C.}\ \bibnamefont {Gou}},
  \bibinfo {author} {\bibfnamefont {I.-K.}\ \bibnamefont {Liu}}, \bibinfo
  {author} {\bibfnamefont {I.~B.}\ \bibnamefont {Spielman}}, \bibinfo {author}
  {\bibfnamefont {L.}~\bibnamefont {Santos}}, \bibinfo {author} {\bibfnamefont
  {A.}~\bibnamefont {Acus}}, \bibinfo {author} {\bibfnamefont {A.}~\bibnamefont
  {Mekys}}, \bibinfo {author} {\bibfnamefont {J.}~\bibnamefont {Ruseckas}}, \
  and\ \bibinfo {author} {\bibfnamefont {G.}~\bibnamefont {Juzeli{\=u}nas}},\
  }\href@noop {} {\bibfield  {journal} {\bibinfo  {journal} {New J. Phys.}\
  }\textbf {\bibinfo {volume} {17}},\ \bibinfo {pages} {033045} (\bibinfo
  {year} {2015})}\BibitemShut {NoStop}%
\bibitem [{\citenamefont {Jiang}\ \emph {et~al.}(2011)\citenamefont {Jiang},
  \citenamefont {Liu}, \citenamefont {Hu},\ and\ \citenamefont
  {Pu}}]{Jiang2011}%
  \BibitemOpen
  \bibfield  {author} {\bibinfo {author} {\bibfnamefont {L.}~\bibnamefont
  {Jiang}}, \bibinfo {author} {\bibfnamefont {X.-J.}\ \bibnamefont {Liu}},
  \bibinfo {author} {\bibfnamefont {H.}~\bibnamefont {Hu}}, \ and\ \bibinfo
  {author} {\bibfnamefont {H.}~\bibnamefont {Pu}},\ }\href@noop {} {\bibfield
  {journal} {\bibinfo  {journal} {Phys. Rev. A}\ }\textbf {\bibinfo {volume}
  {84}},\ \bibinfo {pages} {063618} (\bibinfo {year} {2011})}\BibitemShut
  {NoStop}%
\bibitem [{\citenamefont {Vyasanakere}\ and\ \citenamefont
  {Shenoy}(2012)}]{Vyasanakere2012}%
  \BibitemOpen
  \bibfield  {author} {\bibinfo {author} {\bibfnamefont {J.~P.}\ \bibnamefont
  {Vyasanakere}}\ and\ \bibinfo {author} {\bibfnamefont {V.~B.}\ \bibnamefont
  {Shenoy}},\ }\href@noop {} {\bibfield  {journal} {\bibinfo  {journal} {New J.
  Phys.}\ }\textbf {\bibinfo {volume} {14}},\ \bibinfo {pages} {043041}
  (\bibinfo {year} {2012})}\BibitemShut {NoStop}%
\bibitem [{\citenamefont {Zhou}\ \emph
  {et~al.}(2011{\natexlab{b}})\citenamefont {Zhou}, \citenamefont {Zhang},\
  and\ \citenamefont {Yi}}]{Zhou2011PRA}%
  \BibitemOpen
  \bibfield  {author} {\bibinfo {author} {\bibfnamefont {J.}~\bibnamefont
  {Zhou}}, \bibinfo {author} {\bibfnamefont {W.}~\bibnamefont {Zhang}}, \ and\
  \bibinfo {author} {\bibfnamefont {W.}~\bibnamefont {Yi}},\ }\href@noop {}
  {\bibfield  {journal} {\bibinfo  {journal} {Phys. Rev. A}\ }\textbf {\bibinfo
  {volume} {84}},\ \bibinfo {pages} {063603} (\bibinfo {year}
  {2011}{\natexlab{b}})}\BibitemShut {NoStop}%
\bibitem [{\citenamefont {Liu}\ and\ \citenamefont {Hu}(2012)}]{Liu2012}%
  \BibitemOpen
  \bibfield  {author} {\bibinfo {author} {\bibfnamefont {X.-J.}\ \bibnamefont
  {Liu}}\ and\ \bibinfo {author} {\bibfnamefont {H.}~\bibnamefont {Hu}},\
  }\href@noop {} {\bibfield  {journal} {\bibinfo  {journal} {Phys. Rev. A}\
  }\textbf {\bibinfo {volume} {85}},\ \bibinfo {pages} {033622} (\bibinfo
  {year} {2012})}\BibitemShut {NoStop}%
\bibitem [{\citenamefont {Liu}\ and\ \citenamefont {Hu}(2013)}]{Liu2013}%
  \BibitemOpen
  \bibfield  {author} {\bibinfo {author} {\bibfnamefont {X.-J.}\ \bibnamefont
  {Liu}}\ and\ \bibinfo {author} {\bibfnamefont {H.}~\bibnamefont {Hu}},\
  }\href@noop {} {\bibfield  {journal} {\bibinfo  {journal} {Phys. Rev. A}\
  }\textbf {\bibinfo {volume} {88}},\ \bibinfo {pages} {023622} (\bibinfo
  {year} {2013})}\BibitemShut {NoStop}%
\bibitem [{\citenamefont {Cai}\ \emph {et~al.}(2012)\citenamefont {Cai},
  \citenamefont {Zhou},\ and\ \citenamefont {Wu}}]{Cai2012}%
  \BibitemOpen
  \bibfield  {author} {\bibinfo {author} {\bibfnamefont {Z.}~\bibnamefont
  {Cai}}, \bibinfo {author} {\bibfnamefont {X.}~\bibnamefont {Zhou}}, \ and\
  \bibinfo {author} {\bibfnamefont {C.}~\bibnamefont {Wu}},\ }\href@noop {}
  {\bibfield  {journal} {\bibinfo  {journal} {Phys. Rev. A}\ }\textbf {\bibinfo
  {volume} {85}},\ \bibinfo {pages} {061605(R)} (\bibinfo {year}
  {2012})}\BibitemShut {NoStop}%
\bibitem [{\citenamefont {Radi{\'c}}\ \emph {et~al.}(2012)\citenamefont
  {Radi{\'c}}, \citenamefont {DiCiolo}, \citenamefont {Sun},\ and\
  \citenamefont {Galitski}}]{Radic2012}%
  \BibitemOpen
  \bibfield  {author} {\bibinfo {author} {\bibfnamefont {J.}~\bibnamefont
  {Radi{\'c}}}, \bibinfo {author} {\bibfnamefont {A.}~\bibnamefont {DiCiolo}},
  \bibinfo {author} {\bibfnamefont {K.}~\bibnamefont {Sun}}, \ and\ \bibinfo
  {author} {\bibfnamefont {V.}~\bibnamefont {Galitski}},\ }\href@noop {}
  {\bibfield  {journal} {\bibinfo  {journal} {Phys. Rev. Lett.}\ }\textbf
  {\bibinfo {volume} {109}},\ \bibinfo {pages} {085303} (\bibinfo {year}
  {2012})}\BibitemShut {NoStop}%
\bibitem [{\citenamefont {Cole}\ \emph {et~al.}(2012)\citenamefont {Cole},
  \citenamefont {Zhang}, \citenamefont {Paramekanti},\ and\ \citenamefont
  {Trivedi}}]{Cole2012}%
  \BibitemOpen
  \bibfield  {author} {\bibinfo {author} {\bibfnamefont {W.~S.}\ \bibnamefont
  {Cole}}, \bibinfo {author} {\bibfnamefont {S.}~\bibnamefont {Zhang}},
  \bibinfo {author} {\bibfnamefont {A.}~\bibnamefont {Paramekanti}}, \ and\
  \bibinfo {author} {\bibfnamefont {N.}~\bibnamefont {Trivedi}},\ }\href@noop
  {} {\bibfield  {journal} {\bibinfo  {journal} {Phys. Rev. Lett.}\ }\textbf
  {\bibinfo {volume} {109}},\ \bibinfo {pages} {085302} (\bibinfo {year}
  {2012})}\BibitemShut {NoStop}%
\bibitem [{\citenamefont {Xu}\ \emph {et~al.}(2014)\citenamefont {Xu},
  \citenamefont {Cole},\ and\ \citenamefont {Zhang}}]{Xu2014}%
  \BibitemOpen
  \bibfield  {author} {\bibinfo {author} {\bibfnamefont {Z.}~\bibnamefont
  {Xu}}, \bibinfo {author} {\bibfnamefont {W.~S.}\ \bibnamefont {Cole}}, \ and\
  \bibinfo {author} {\bibfnamefont {S.}~\bibnamefont {Zhang}},\ }\href@noop {}
  {\bibfield  {journal} {\bibinfo  {journal} {Phys. Rev. A}\ }\textbf {\bibinfo
  {volume} {89}},\ \bibinfo {pages} {051604(R)} (\bibinfo {year}
  {2014})}\BibitemShut {NoStop}%
\bibitem [{\citenamefont {Chen}\ and\ \citenamefont {Liang}(2016)}]{Chen2016}%
  \BibitemOpen
  \bibfield  {author} {\bibinfo {author} {\bibfnamefont {Z.}~\bibnamefont
  {Chen}}\ and\ \bibinfo {author} {\bibfnamefont {Z.}~\bibnamefont {Liang}},\
  }\href@noop {} {\bibfield  {journal} {\bibinfo  {journal} {Phys. Rev. A}\
  }\textbf {\bibinfo {volume} {93}},\ \bibinfo {pages} {013601} (\bibinfo
  {year} {2016})}\BibitemShut {NoStop}%
\bibitem [{\citenamefont {Lin}\ \emph {et~al.}(2011)\citenamefont {Lin},
  \citenamefont {Jim\'{e}nez-Garc\'{i}a},\ and\ \citenamefont
  {Spielman}}]{Lin2011}%
  \BibitemOpen
  \bibfield  {author} {\bibinfo {author} {\bibfnamefont {Y.-J.}\ \bibnamefont
  {Lin}}, \bibinfo {author} {\bibfnamefont {K.}~\bibnamefont
  {Jim\'{e}nez-Garc\'{i}a}}, \ and\ \bibinfo {author} {\bibfnamefont {I.~B.}\
  \bibnamefont {Spielman}},\ }\href@noop {} {\bibfield  {journal} {\bibinfo
  {journal} {Nature}\ }\textbf {\bibinfo {volume} {471}},\ \bibinfo {pages}
  {83} (\bibinfo {year} {2011})}\BibitemShut {NoStop}%
\bibitem [{\citenamefont {Zhang}\ \emph {et~al.}(2012)\citenamefont {Zhang},
  \citenamefont {Ji}, \citenamefont {Chen}, \citenamefont {Zhang},
  \citenamefont {Du}, \citenamefont {Yan}, \citenamefont {Pan}, \citenamefont
  {Zhao}, \citenamefont {Deng}, \citenamefont {Zhai}, \citenamefont {Chen},\
  and\ \citenamefont {Pan}}]{Zhang2012}%
  \BibitemOpen
  \bibfield  {author} {\bibinfo {author} {\bibfnamefont {J.-Y.}\ \bibnamefont
  {Zhang}}, \bibinfo {author} {\bibfnamefont {S.-C.}\ \bibnamefont {Ji}},
  \bibinfo {author} {\bibfnamefont {Z.}~\bibnamefont {Chen}}, \bibinfo {author}
  {\bibfnamefont {L.}~\bibnamefont {Zhang}}, \bibinfo {author} {\bibfnamefont
  {Z.-D.}\ \bibnamefont {Du}}, \bibinfo {author} {\bibfnamefont
  {B.}~\bibnamefont {Yan}}, \bibinfo {author} {\bibfnamefont {G.-S.}\
  \bibnamefont {Pan}}, \bibinfo {author} {\bibfnamefont {B.}~\bibnamefont
  {Zhao}}, \bibinfo {author} {\bibfnamefont {Y.-J.}\ \bibnamefont {Deng}},
  \bibinfo {author} {\bibfnamefont {H.}~\bibnamefont {Zhai}}, \bibinfo {author}
  {\bibfnamefont {S.}~\bibnamefont {Chen}}, \ and\ \bibinfo {author}
  {\bibfnamefont {J.-W.}\ \bibnamefont {Pan}},\ }\href@noop {} {\bibfield
  {journal} {\bibinfo  {journal} {Phys. Rev. Lett.}\ }\textbf {\bibinfo
  {volume} {109}},\ \bibinfo {pages} {115301} (\bibinfo {year}
  {2012})}\BibitemShut {NoStop}%
\bibitem [{\citenamefont {Wang}\ \emph {et~al.}(2012)\citenamefont {Wang},
  \citenamefont {Yu}, \citenamefont {Fu}, \citenamefont {Miao}, \citenamefont
  {Huang}, \citenamefont {Chai}, \citenamefont {Zhai},\ and\ \citenamefont
  {Zhang}}]{Wang2012}%
  \BibitemOpen
  \bibfield  {author} {\bibinfo {author} {\bibfnamefont {P.}~\bibnamefont
  {Wang}}, \bibinfo {author} {\bibfnamefont {Z.-Q.}\ \bibnamefont {Yu}},
  \bibinfo {author} {\bibfnamefont {Z.}~\bibnamefont {Fu}}, \bibinfo {author}
  {\bibfnamefont {J.}~\bibnamefont {Miao}}, \bibinfo {author} {\bibfnamefont
  {L.}~\bibnamefont {Huang}}, \bibinfo {author} {\bibfnamefont
  {S.}~\bibnamefont {Chai}}, \bibinfo {author} {\bibfnamefont {H.}~\bibnamefont
  {Zhai}}, \ and\ \bibinfo {author} {\bibfnamefont {J.}~\bibnamefont {Zhang}},\
  }\href@noop {} {\bibfield  {journal} {\bibinfo  {journal} {Phys. Rev. Lett.}\
  }\textbf {\bibinfo {volume} {109}},\ \bibinfo {pages} {095301} (\bibinfo
  {year} {2012})}\BibitemShut {NoStop}%
\bibitem [{\citenamefont {Cheuk}\ \emph {et~al.}(2012)\citenamefont {Cheuk},
  \citenamefont {Sommer}, \citenamefont {Hadzibabic}, \citenamefont {Yefsah},
  \citenamefont {Bakr},\ and\ \citenamefont {Zwierlein}}]{Cheuk2012}%
  \BibitemOpen
  \bibfield  {author} {\bibinfo {author} {\bibfnamefont {L.~W.}\ \bibnamefont
  {Cheuk}}, \bibinfo {author} {\bibfnamefont {A.~T.}\ \bibnamefont {Sommer}},
  \bibinfo {author} {\bibfnamefont {Z.}~\bibnamefont {Hadzibabic}}, \bibinfo
  {author} {\bibfnamefont {T.}~\bibnamefont {Yefsah}}, \bibinfo {author}
  {\bibfnamefont {W.~S.}\ \bibnamefont {Bakr}}, \ and\ \bibinfo {author}
  {\bibfnamefont {M.~W.}\ \bibnamefont {Zwierlein}},\ }\href@noop {} {\bibfield
   {journal} {\bibinfo  {journal} {Phys. Rev. Lett.}\ }\textbf {\bibinfo
  {volume} {109}},\ \bibinfo {pages} {095302} (\bibinfo {year}
  {2012})}\BibitemShut {NoStop}%
\bibitem [{\citenamefont {Williams}\ \emph {et~al.}(2012)\citenamefont
  {Williams}, \citenamefont {LeBlanc}, \citenamefont {Jim\'enez-Garc\'\i{}a},
  \citenamefont {Beeler}, \citenamefont {Perry}, \citenamefont {Phillips},\
  and\ \citenamefont {Spielman}}]{Williams2012}%
  \BibitemOpen
  \bibfield  {author} {\bibinfo {author} {\bibfnamefont {R.~A.}\ \bibnamefont
  {Williams}}, \bibinfo {author} {\bibfnamefont {L.~J.}\ \bibnamefont
  {LeBlanc}}, \bibinfo {author} {\bibfnamefont {K.}~\bibnamefont
  {Jim\'enez-Garc\'\i{}a}}, \bibinfo {author} {\bibfnamefont {M.~C.}\
  \bibnamefont {Beeler}}, \bibinfo {author} {\bibfnamefont {A.~R.}\
  \bibnamefont {Perry}}, \bibinfo {author} {\bibfnamefont {W.~D.}\ \bibnamefont
  {Phillips}}, \ and\ \bibinfo {author} {\bibfnamefont {I.~B.}\ \bibnamefont
  {Spielman}},\ }\href@noop {} {\bibfield  {journal} {\bibinfo  {journal}
  {Science}\ }\textbf {\bibinfo {volume} {335}},\ \bibinfo {pages} {314}
  (\bibinfo {year} {2012})}\BibitemShut {NoStop}%
\bibitem [{\citenamefont {LeBlanc}\ \emph {et~al.}(2013)\citenamefont
  {LeBlanc}, \citenamefont {Beeler}, \citenamefont {Jimenez-Garcia},
  \citenamefont {Perry}, \citenamefont {Sugawa}, \citenamefont {Williams},\
  and\ \citenamefont {Spielman}}]{LeBlanc2013}%
  \BibitemOpen
  \bibfield  {author} {\bibinfo {author} {\bibfnamefont {L.~J.}\ \bibnamefont
  {LeBlanc}}, \bibinfo {author} {\bibfnamefont {M.~C.}\ \bibnamefont {Beeler}},
  \bibinfo {author} {\bibfnamefont {K.}~\bibnamefont {Jimenez-Garcia}},
  \bibinfo {author} {\bibfnamefont {A.~R.}\ \bibnamefont {Perry}}, \bibinfo
  {author} {\bibfnamefont {S.}~\bibnamefont {Sugawa}}, \bibinfo {author}
  {\bibfnamefont {R.~A.}\ \bibnamefont {Williams}}, \ and\ \bibinfo {author}
  {\bibfnamefont {I.~B.}\ \bibnamefont {Spielman}},\ }\href@noop {} {\bibfield
  {journal} {\bibinfo  {journal} {New. J. Phys.}\ }\textbf {\bibinfo {volume}
  {15}},\ \bibinfo {pages} {073011} (\bibinfo {year} {2013})}\BibitemShut
  {NoStop}%
\bibitem [{\citenamefont {Qu}\ \emph {et~al.}(2013)\citenamefont {Qu},
  \citenamefont {Hamner}, \citenamefont {Gong}, \citenamefont {Zhang},\ and\
  \citenamefont {Engels}}]{Engels2013}%
  \BibitemOpen
  \bibfield  {author} {\bibinfo {author} {\bibfnamefont {C.}~\bibnamefont
  {Qu}}, \bibinfo {author} {\bibfnamefont {C.}~\bibnamefont {Hamner}}, \bibinfo
  {author} {\bibfnamefont {M.}~\bibnamefont {Gong}}, \bibinfo {author}
  {\bibfnamefont {C.}~\bibnamefont {Zhang}}, \ and\ \bibinfo {author}
  {\bibfnamefont {P.}~\bibnamefont {Engels}},\ }\href@noop {} {\bibfield
  {journal} {\bibinfo  {journal} {Phys. Rev. A}\ }\textbf {\bibinfo {volume}
  {88}},\ \bibinfo {pages} {021604(R)} (\bibinfo {year} {2013})}\BibitemShut
  {NoStop}%
\bibitem [{\citenamefont {Fu}\ \emph {et~al.}(2014)\citenamefont {Fu},
  \citenamefont {Huang}, \citenamefont {Meng}, \citenamefont {Wang},
  \citenamefont {Zhang}, \citenamefont {Zhang}, \citenamefont {Zhai},
  \citenamefont {Zhang},\ and\ \citenamefont {Zhang}}]{Fu2014}%
  \BibitemOpen
  \bibfield  {author} {\bibinfo {author} {\bibfnamefont {Z.}~\bibnamefont
  {Fu}}, \bibinfo {author} {\bibfnamefont {L.}~\bibnamefont {Huang}}, \bibinfo
  {author} {\bibfnamefont {Z.}~\bibnamefont {Meng}}, \bibinfo {author}
  {\bibfnamefont {P.}~\bibnamefont {Wang}}, \bibinfo {author} {\bibfnamefont
  {L.}~\bibnamefont {Zhang}}, \bibinfo {author} {\bibfnamefont
  {S.}~\bibnamefont {Zhang}}, \bibinfo {author} {\bibfnamefont
  {H.}~\bibnamefont {Zhai}}, \bibinfo {author} {\bibfnamefont {P.}~\bibnamefont
  {Zhang}}, \ and\ \bibinfo {author} {\bibfnamefont {J.}~\bibnamefont
  {Zhang}},\ }\href@noop {} {\bibfield  {journal} {\bibinfo  {journal} {Nature
  Phys.}\ }\textbf {\bibinfo {volume} {10}}, \ \bibinfo {pages} {815} (\bibinfo {year}
  {2014})}\BibitemShut {NoStop}%
\bibitem [{\citenamefont {Liu}\ \emph {et~al.}(2009)\citenamefont {Liu},
  \citenamefont {Borunda}, \citenamefont {Liu},\ and\ \citenamefont
  {Sinova}}]{Liu2009}%
  \BibitemOpen
  \bibfield  {author} {\bibinfo {author} {\bibfnamefont {X.-J.}\ \bibnamefont
  {Liu}}, \bibinfo {author} {\bibfnamefont {M.~F.}\ \bibnamefont {Borunda}},
  \bibinfo {author} {\bibfnamefont {X.}~\bibnamefont {Liu}}, \ and\ \bibinfo
  {author} {\bibfnamefont {J.}~\bibnamefont {Sinova}},\ }\href@noop {}
  {\bibfield  {journal} {\bibinfo  {journal} {Phys. Rev. Lett.}\ }\textbf
  {\bibinfo {volume} {102}},\ \bibinfo {pages} {046402} (\bibinfo {year}
  {2009})}\BibitemShut {NoStop}%
\bibitem [{\citenamefont {Li}\ \emph {et~al.}(2012)\citenamefont {Li},
  \citenamefont {Pitaevskii},\ and\ \citenamefont {Stringari}}]{Li2012}%
  \BibitemOpen
  \bibfield  {author} {\bibinfo {author} {\bibfnamefont {Y.}~\bibnamefont
  {Li}}, \bibinfo {author} {\bibfnamefont {L.~P.}\ \bibnamefont {Pitaevskii}},
  \ and\ \bibinfo {author} {\bibfnamefont {S.}~\bibnamefont {Stringari}},\
  }\href@noop {} {\bibfield  {journal} {\bibinfo  {journal} {Phys. Rev. Lett.}\
  }\textbf {\bibinfo {volume} {108}},\ \bibinfo {pages} {225301} (\bibinfo
  {year} {2012})}\BibitemShut {NoStop}%
\bibitem [{\citenamefont {Li}\ \emph {et~al.}(2013)\citenamefont {Li},
  \citenamefont {Martone}, \citenamefont {Pitaevskii},\ and\ \citenamefont
  {Stringari}}]{Li2013}%
  \BibitemOpen
  \bibfield  {author} {\bibinfo {author} {\bibfnamefont {Y.}~\bibnamefont
  {Li}}, \bibinfo {author} {\bibfnamefont {G.~I.}\ \bibnamefont {Martone}},
  \bibinfo {author} {\bibfnamefont {L.~P.}\ \bibnamefont {Pitaevskii}}, \ and\
  \bibinfo {author} {\bibfnamefont {S.}~\bibnamefont {Stringari}},\ }\href@noop
  {} {\bibfield  {journal} {\bibinfo  {journal} {Phys. Rev. Lett.}\ }\textbf
  {\bibinfo {volume} {110}},\ \bibinfo {pages} {235302} (\bibinfo {year}
  {2013})}\BibitemShut {NoStop}%
\bibitem [{\citenamefont {Stanescu}\ \emph {et~al.}(2007)\citenamefont
  {Stanescu}, \citenamefont {Zhang},\ and\ \citenamefont
  {Galitski}}]{Stanescu2007}%
  \BibitemOpen
  \bibfield  {author} {\bibinfo {author} {\bibfnamefont {T.~D.}\ \bibnamefont
  {Stanescu}}, \bibinfo {author} {\bibfnamefont {C.}~\bibnamefont {Zhang}}, \
  and\ \bibinfo {author} {\bibfnamefont {V.}~\bibnamefont {Galitski}},\
  }\href@noop {} {\bibfield  {journal} {\bibinfo  {journal} {Phys. Rev. Lett.}\
  }\textbf {\bibinfo {volume} {99}},\ \bibinfo {pages} {110403} (\bibinfo
  {year} {2007})}\BibitemShut {NoStop}%
\bibitem [{\citenamefont {Jacob}\ \emph {et~al.}(2007)\citenamefont {Jacob},
  \citenamefont {{\"O}hberg}, \citenamefont {Juzeli{\=u}nas},\ and\
  \citenamefont {Santos}}]{Jacob2007}%
  \BibitemOpen
  \bibfield  {author} {\bibinfo {author} {\bibfnamefont {A.}~\bibnamefont
  {Jacob}}, \bibinfo {author} {\bibfnamefont {P.}~\bibnamefont {{\"O}hberg}},
  \bibinfo {author} {\bibfnamefont {G.}~\bibnamefont {Juzeli{\=u}nas}}, \ and\
  \bibinfo {author} {\bibfnamefont {L.}~\bibnamefont {Santos}},\ }\href@noop {}
  {\bibfield  {journal} {\bibinfo  {journal} {Appl. Phys. B}\ }\textbf
  {\bibinfo {volume} {89}},\ \bibinfo {pages} {439} (\bibinfo {year}
  {2007})}\BibitemShut {NoStop}%
\bibitem [{\citenamefont {Stanescu}\ \emph {et~al.}(2008)\citenamefont
  {Stanescu}, \citenamefont {Anderson},\ and\ \citenamefont
  {Galitski}}]{Stanescu2008}%
  \BibitemOpen
  \bibfield  {author} {\bibinfo {author} {\bibfnamefont {T.~D.}\ \bibnamefont
  {Stanescu}}, \bibinfo {author} {\bibfnamefont {B.}~\bibnamefont {Anderson}},
  \ and\ \bibinfo {author} {\bibfnamefont {V.}~\bibnamefont {Galitski}},\
  }\href@noop {} {\bibfield  {journal} {\bibinfo  {journal} {Phys. Rev. A}\
  }\textbf {\bibinfo {volume} {78}},\ \bibinfo {pages} {023616} (\bibinfo
  {year} {2008})}\BibitemShut {NoStop}%
\bibitem [{\citenamefont {Juzeli{\=u}nas}\ \emph {et~al.}(2008)\citenamefont
  {Juzeli{\=u}nas}, \citenamefont {Ruseckas}, \citenamefont {Lindberg},
  \citenamefont {Santos},\ and\ \citenamefont
  {{\"O}hberg}}]{Juzeliunas2008PRA}%
  \BibitemOpen
  \bibfield  {author} {\bibinfo {author} {\bibfnamefont {G.}~\bibnamefont
  {Juzeli{\=u}nas}}, \bibinfo {author} {\bibfnamefont {J.}~\bibnamefont
  {Ruseckas}}, \bibinfo {author} {\bibfnamefont {M.}~\bibnamefont {Lindberg}},
  \bibinfo {author} {\bibfnamefont {L.}~\bibnamefont {Santos}}, \ and\ \bibinfo
  {author} {\bibfnamefont {P.}~\bibnamefont {{\"O}hberg}},\ }\href@noop {}
  {\bibfield  {journal} {\bibinfo  {journal} {Phys. Rev. A}\ }\textbf {\bibinfo
  {volume} {77}},\ \bibinfo {pages} {011802(R)} (\bibinfo {year}
  {2008})}\BibitemShut {NoStop}%
\bibitem [{\citenamefont {Zhang}(2010)}]{Chuanwei-Zhang10}%
  \BibitemOpen
  \bibfield  {author} {\bibinfo {author} {\bibfnamefont {C.}~\bibnamefont
  {Zhang}},\ }\href@noop {} {\bibfield  {journal} {\bibinfo  {journal} {Phys.
  Rev. A}\ }\textbf {\bibinfo {volume} {82}},\ \bibinfo {pages} {021607(R)}
  (\bibinfo {year} {2010})}\BibitemShut {NoStop}%
\bibitem [{\citenamefont {Campbell}\ \emph {et~al.}(2011)\citenamefont
  {Campbell}, \citenamefont {Juzeli{\=u}nas},\ and\ \citenamefont
  {Spielman}}]{Campbell2011}%
  \BibitemOpen
  \bibfield  {author} {\bibinfo {author} {\bibfnamefont {D.~L.}\ \bibnamefont
  {Campbell}}, \bibinfo {author} {\bibfnamefont {G.}~\bibnamefont
  {Juzeli{\=u}nas}}, \ and\ \bibinfo {author} {\bibfnamefont {I.~B.}\
  \bibnamefont {Spielman}},\ }\href@noop {} {\bibfield  {journal} {\bibinfo
  {journal} {Phys. Rev. A}\ }\textbf {\bibinfo {volume} {84}},\ \bibinfo
  {pages} {025602} (\bibinfo {year} {2011})}\BibitemShut {NoStop}%
\bibitem [{\citenamefont {Su}\ \emph {et~al.}(2012)\citenamefont {Su},
  \citenamefont {Liu}, \citenamefont {Tsai}, \citenamefont {Liu},\ and\
  \citenamefont {Gou}}]{Su2012}%
  \BibitemOpen
  \bibfield  {author} {\bibinfo {author} {\bibfnamefont {S.-W.}\ \bibnamefont
  {Su}}, \bibinfo {author} {\bibfnamefont {I.-K.}\ \bibnamefont {Liu}},
  \bibinfo {author} {\bibfnamefont {Y.-C.}\ \bibnamefont {Tsai}}, \bibinfo
  {author} {\bibfnamefont {W.-M.}\ \bibnamefont {Liu}}, \ and\ \bibinfo
  {author} {\bibfnamefont {S.-C.}\ \bibnamefont {Gou}},\ }\href@noop {}
  {\bibfield  {journal} {\bibinfo  {journal} {Phys. Rev. A}\ }\textbf {\bibinfo
  {volume} {86}},\ \bibinfo {pages} {023601} (\bibinfo {year}
  {2012})}\BibitemShut {NoStop}%
\bibitem [{\citenamefont {Anderson}\ \emph {et~al.}(2012)\citenamefont
  {Anderson}, \citenamefont {Juzeli{\=u}nas}, \citenamefont {Galitski},\ and\
  \citenamefont {Spielman}}]{Anderson2012PRL}%
  \BibitemOpen
  \bibfield  {author} {\bibinfo {author} {\bibfnamefont {B.~M.}\ \bibnamefont
  {Anderson}}, \bibinfo {author} {\bibfnamefont {G.}~\bibnamefont
  {Juzeli{\=u}nas}}, \bibinfo {author} {\bibfnamefont {V.~M.}\ \bibnamefont
  {Galitski}}, \ and\ \bibinfo {author} {\bibfnamefont {I.~B.}\ \bibnamefont
  {Spielman}},\ }\href@noop {} {\bibfield  {journal} {\bibinfo  {journal}
  {Phys. Rev. Lett.}\ }\textbf {\bibinfo {volume} {108}},\ \bibinfo {pages}
  {235301} (\bibinfo {year} {2012})}\BibitemShut {NoStop}%
\bibitem [{\citenamefont {Xu}\ \emph {et~al.}(2013)\citenamefont {Xu},
  \citenamefont {You},\ and\ \citenamefont {Ueda}}]{Xu2013}%
  \BibitemOpen
  \bibfield  {author} {\bibinfo {author} {\bibfnamefont {Z.-F.}\ \bibnamefont
  {Xu}}, \bibinfo {author} {\bibfnamefont {L.}~\bibnamefont {You}}, \ and\
  \bibinfo {author} {\bibfnamefont {M.}~\bibnamefont {Ueda}},\ }\href@noop {}
  {\bibfield  {journal} {\bibinfo  {journal} {Phys. Rev. A}\ }\textbf {\bibinfo
  {volume} {87}},\ \bibinfo {pages} {063634} (\bibinfo {year}
  {2013})}\BibitemShut {NoStop}%
\bibitem [{\citenamefont {Anderson}\ \emph {et~al.}(2013)\citenamefont
  {Anderson}, \citenamefont {Spielman},\ and\ \citenamefont
  {Juzeli{\=u}nas}}]{Anderson2013}%
  \BibitemOpen
  \bibfield  {author} {\bibinfo {author} {\bibfnamefont {B.~M.}\ \bibnamefont
  {Anderson}}, \bibinfo {author} {\bibfnamefont {I.~B.}\ \bibnamefont
  {Spielman}}, \ and\ \bibinfo {author} {\bibfnamefont {G.}~\bibnamefont
  {Juzeli{\=u}nas}},\ }\href@noop {} {\bibfield  {journal} {\bibinfo  {journal}
  {Phys. Rev. Lett.}\ }\textbf {\bibinfo {volume} {111}},\ \bibinfo {pages}
  {125301} (\bibinfo {year} {2013})}\BibitemShut {NoStop}%
\bibitem [{\citenamefont {Sun}\ \emph {et~al.}(2015)\citenamefont {Sun},
  \citenamefont {Wen}, \citenamefont {Liu}, \citenamefont {Juzeli{\=u}nas},\
  and\ \citenamefont {Ji}}]{Sun2015PRA}%
  \BibitemOpen
  \bibfield  {author} {\bibinfo {author} {\bibfnamefont {Q.}~\bibnamefont
  {Sun}}, \bibinfo {author} {\bibfnamefont {L.}~\bibnamefont {Wen}}, \bibinfo
  {author} {\bibfnamefont {W.-M.}\ \bibnamefont {Liu}}, \bibinfo {author}
  {\bibfnamefont {G.}~\bibnamefont {Juzeli{\=u}nas}}, \ and\ \bibinfo {author}
  {\bibfnamefont {A.-C.}\ \bibnamefont {Ji}},\ }\href@noop {} {\bibfield
  {journal} {\bibinfo  {journal} {Phys. Rev. A}\ }\textbf {\bibinfo {volume}
  {91}},\ \bibinfo {pages} {033619} (\bibinfo {year} {2015})}\BibitemShut
  {NoStop}%
\bibitem [{\citenamefont {Huang}\ \emph {et~al.}(2016)\citenamefont {Huang},
  \citenamefont {Meng}, \citenamefont {Wang}, \citenamefont {Peng},
  \citenamefont {Zhang}, \citenamefont {Chen}, \citenamefont {Li},
  \citenamefont {Zhou},\ and\ \citenamefont {Zhang}}]{Huang2016}%
  \BibitemOpen
  \bibfield  {author} {\bibinfo {author} {\bibfnamefont {L.}~\bibnamefont
  {Huang}}, \bibinfo {author} {\bibfnamefont {Z.}~\bibnamefont {Meng}},
  \bibinfo {author} {\bibfnamefont {P.}~\bibnamefont {Wang}}, \bibinfo {author}
  {\bibfnamefont {P.}~\bibnamefont {Peng}}, \bibinfo {author} {\bibfnamefont
  {S.-L.}\ \bibnamefont {Zhang}}, \bibinfo {author} {\bibfnamefont
  {L.}~\bibnamefont {Chen}}, \bibinfo {author} {\bibfnamefont {D.}~\bibnamefont
  {Li}}, \bibinfo {author} {\bibfnamefont {Q.}~\bibnamefont {Zhou}}, \ and\
  \bibinfo {author} {\bibfnamefont {J.}~\bibnamefont {Zhang}},\ }\href@noop {}
  {\bibfield  {journal} {\bibinfo  {journal} {Nature Phys.}\ }\textbf {\bibinfo
  {volume} {doi:10.1038/nphys3672}} (\bibinfo {year} {2016})}\BibitemShut
  {NoStop}%
\bibitem [{\citenamefont {Meng}\ \emph {et~al.}(2015)\citenamefont {Meng},
  \citenamefont {Huang}, \citenamefont {Peng}, \citenamefont {Li},
  \citenamefont {Chen}, \citenamefont {Xu}, \citenamefont {Zhang},
  \citenamefont {Wang},\ and\ \citenamefont {Zhang}}]{Meng2015}%
  \BibitemOpen
  \bibfield  {author} {\bibinfo {author} {\bibfnamefont {Z.}~\bibnamefont
  {Meng}}, \bibinfo {author} {\bibfnamefont {L.}~\bibnamefont {Huang}},
  \bibinfo {author} {\bibfnamefont {P.}~\bibnamefont {Peng}}, \bibinfo {author}
  {\bibfnamefont {D.}~\bibnamefont {Li}}, \bibinfo {author} {\bibfnamefont
  {L.}~\bibnamefont {Chen}}, \bibinfo {author} {\bibfnamefont {Y.}~\bibnamefont
  {Xu}}, \bibinfo {author} {\bibfnamefont {C.}~\bibnamefont {Zhang}}, \bibinfo
  {author} {\bibfnamefont {P.}~\bibnamefont {Wang}}, \ and\ \bibinfo {author}
  {\bibfnamefont {J.}~\bibnamefont {Zhang}},\ }\href@noop {} {\bibfield
  {journal} {\bibinfo  {journal} {arXiv:1511.08492}\ } (\bibinfo {year}
  {2015})}\BibitemShut {NoStop}%
\bibitem [{Note1()}]{Note1}%
  \BibitemOpen
  \bibinfo {note} {Two dimensional spin-orbit coupling has also been recently
  realized using another approach which relies on optical lattices, see Z.~Wu,
  L.~Zhang, W.~Sun, X.-T.~Xu, B.-Z.~Wang, S.-C.~Ji, Y.~Deng, S.~Chen, X.-J.~Liu
  and J.-W.~Pan, arXiv:1511.08170.}\BibitemShut {Stop}%
\bibitem [{\citenamefont {Hung}\ \emph {et~al.}(2015)\citenamefont {Hung},
  \citenamefont {Huang}, \citenamefont {Wu}, \citenamefont {Bruvelis},
  \citenamefont {Xiao}, \citenamefont {Ekers},\ and\ \citenamefont
  {Yu}}]{Hung2015}%
  \BibitemOpen
  \bibfield  {author} {\bibinfo {author} {\bibfnamefont {W.}~\bibnamefont
  {Hung}}, \bibinfo {author} {\bibfnamefont {P.}~\bibnamefont {Huang}},
  \bibinfo {author} {\bibfnamefont {F.-C.}\ \bibnamefont {Wu}}, \bibinfo
  {author} {\bibfnamefont {M.}~\bibnamefont {Bruvelis}}, \bibinfo {author}
  {\bibfnamefont {H.-Y.}\ \bibnamefont {Xiao}}, \bibinfo {author}
  {\bibfnamefont {A.}~\bibnamefont {Ekers}}, \ and\ \bibinfo {author}
  {\bibfnamefont {I.~A.}\ \bibnamefont {Yu}},\ }\href@noop {} {\bibfield
  {journal} {\bibinfo  {journal} {J. Opt. Soc. Am. B}\ }\textbf {\bibinfo
  {volume} {32}},\ \bibinfo {pages} {B32} (\bibinfo {year} {2015})}\BibitemShut
  {NoStop}%
\bibitem [{\citenamefont {Sebby-Strabley}\ \emph {et~al.}(2006)\citenamefont
  {Sebby-Strabley}, \citenamefont {Anderlini}, \citenamefont {Jessen},\ and\
  \citenamefont {Porto}}]{Sebby-Strabley2006}%
  \BibitemOpen
  \bibfield  {author} {\bibinfo {author} {\bibfnamefont {J.}~\bibnamefont
  {Sebby-Strabley}}, \bibinfo {author} {\bibfnamefont {M.}~\bibnamefont
  {Anderlini}}, \bibinfo {author} {\bibfnamefont {P.~S.}\ \bibnamefont
  {Jessen}}, \ and\ \bibinfo {author} {\bibfnamefont {J.~V.}\ \bibnamefont
  {Porto}},\ }\href@noop {} {\bibfield  {journal} {\bibinfo  {journal} {Phys.
  Rev. A}\ }\textbf {\bibinfo {volume} {73}},\ \bibinfo {pages} {033605}
  (\bibinfo {year} {2006})}\BibitemShut {NoStop}%
\bibitem [{\citenamefont {Kasamatsu}\ \emph {et~al.}(2005)\citenamefont
  {Kasamatsu}, \citenamefont {Tsubota},\ and\ \citenamefont
  {Ueda}}]{Kasamatsu2005}%
  \BibitemOpen
  \bibfield  {author} {\bibinfo {author} {\bibfnamefont {K.}~\bibnamefont
  {Kasamatsu}}, \bibinfo {author} {\bibfnamefont {M.}~\bibnamefont {Tsubota}},
  \ and\ \bibinfo {author} {\bibfnamefont {M.}~\bibnamefont {Ueda}},\
  }\href@noop {} {\bibfield  {journal} {\bibinfo  {journal} {Phys. Rev. A}\
  }\textbf {\bibinfo {volume} {71}},\ \bibinfo {pages} {043611} (\bibinfo
  {year} {2005})}\BibitemShut {NoStop}%
\bibitem [{\citenamefont {Chin}\ and\ \citenamefont
  {Krotscheck}(2005)}]{Chin2005}%
  \BibitemOpen
  \bibfield  {author} {\bibinfo {author} {\bibfnamefont {S.~A.}\ \bibnamefont
  {Chin}}\ and\ \bibinfo {author} {\bibfnamefont {E.}~\bibnamefont
  {Krotscheck}},\ }\href@noop {} {\bibfield  {journal} {\bibinfo  {journal}
  {Phys. Rev. E}\ }\textbf {\bibinfo {volume} {72}},\ \bibinfo {pages} {036705}
  (\bibinfo {year} {2005})}\BibitemShut {NoStop}%
\bibitem [{\citenamefont {Mermin}\ and\ \citenamefont {Ho}(1976)}]{Mermin1976}%
  \BibitemOpen
  \bibfield  {author} {\bibinfo {author} {\bibfnamefont {N.~D.}\ \bibnamefont
  {Mermin}}\ and\ \bibinfo {author} {\bibfnamefont {T.-L.}\ \bibnamefont
  {Ho}},\ }\href@noop {} {\bibfield  {journal} {\bibinfo  {journal} {Phys. Rev.
  Lett.}\ }\textbf {\bibinfo {volume} {36}},\ \bibinfo {pages} {594} (\bibinfo
  {year} {1976})}\BibitemShut {NoStop}%
\bibitem [{\citenamefont {Mizushima}\ \emph {et~al.}(2002)\citenamefont
  {Mizushima}, \citenamefont {Machida},\ and\ \citenamefont
  {Kita}}]{Mizushima2002}%
  \BibitemOpen
  \bibfield  {author} {\bibinfo {author} {\bibfnamefont {T.}~\bibnamefont
  {Mizushima}}, \bibinfo {author} {\bibfnamefont {K.}~\bibnamefont {Machida}},
  \ and\ \bibinfo {author} {\bibfnamefont {T.}~\bibnamefont {Kita}},\
  }\href@noop {} {\bibfield  {journal} {\bibinfo  {journal} {Phys. Rev. Lett.}\
  }\textbf {\bibinfo {volume} {89}},\ \bibinfo {pages} {030401} (\bibinfo
  {year} {2002})}\BibitemShut {NoStop}%
\bibitem [{\citenamefont {Ho}(1998)}]{Ho1998}%
  \BibitemOpen
  \bibfield  {author} {\bibinfo {author} {\bibfnamefont {T.-L.}\ \bibnamefont
  {Ho}},\ }\href@noop {} {\bibfield  {journal} {\bibinfo  {journal} {Phys. Rev.
  Lett.}\ }\textbf {\bibinfo {volume} {81}},\ \bibinfo {pages} {742} (\bibinfo
  {year} {1998})}\BibitemShut {NoStop}%
\bibitem [{\citenamefont {Ohmi}\ and\ \citenamefont
  {Machida}(1998)}]{Ohmi1998}%
  \BibitemOpen
  \bibfield  {author} {\bibinfo {author} {\bibfnamefont {T.}~\bibnamefont
  {Ohmi}}\ and\ \bibinfo {author} {\bibfnamefont {K.}~\bibnamefont {Machida}},\
  }\href@noop {} {\bibfield  {journal} {\bibinfo  {journal} {J. Phys. Soc.
  Jpn.}\ }\textbf {\bibinfo {volume} {67}},\ \bibinfo {pages} {1822} (\bibinfo
  {year} {1998})}\BibitemShut {NoStop}%
\bibitem [{\citenamefont {Deutsch}\ and\ \citenamefont
  {Jessen}(1998)}]{Deutsch1998}%
  \BibitemOpen
  \bibfield  {author} {\bibinfo {author} {\bibfnamefont {I.~H.}\ \bibnamefont
  {Deutsch}}\ and\ \bibinfo {author} {\bibfnamefont {P.~S.}\ \bibnamefont
  {Jessen}},\ }\href@noop {} {\bibfield  {journal} {\bibinfo  {journal} {Phys.
  Rev. A}\ }\textbf {\bibinfo {volume} {57}},\ \bibinfo {pages} {1972}
  (\bibinfo {year} {1998})}\BibitemShut {NoStop}%
\end{thebibliography}

\end{document}